\documentclass[3p,preprintnumbers]{elsarticle}

\usepackage{lineno,hyperref}
\usepackage{color}
\usepackage{pifont} 
\usepackage{amssymb} 
\newcommand{\md}{\mathrm{d}}
\newcommand{\Ltd}{\Lambda_{\rm 3D}}
\newcommand{\Lfd}{\Lambda_{\rm 4D}}
\newcommand{\Lpv}{\Lambda_{\rm PV}}
\newcommand{\Lpt}{\Lambda_{\rm PT}}

\begin{document}

\begin{frontmatter}

\begin{flushright}
  \pprinttitle{HUPD-1511}
\end{flushright}

\title{Parameter fitting in three-flavor Nambu--Jona-Lasinio model\\
with various regularizations}

\author{H. Kohyama}
\address{Department of Physics,
National Taiwan University, Taipei 10617, Taiwan}

\author{D. Kimura}
\address{General Education, Ube National College of Technology,
Ube, Yamaguchi 755-8555, Japan}

\author{T. Inagaki}
\address{
Information Media Center, Hiroshima University, Higashi-Hiroshima,
Hiroshima 739-8521, Japan,
Core of Research for the Energetic Universe, Hiroshima University,
Higashi-Hiroshima, Japan 739-8526
}




\begin{abstract}
We study the three-flavor Nambu--Jona-Lasinio model with various
regularization procedures.  
We perform parameter fitting in each regularization and apply the obtained
parameter sets to evaluate various physical quantities, several light meson
masses, decay constant and the topological susceptibility. The model
parameters are adopted even at very high cutoff scale compare to the
hadronic scale to study the asymptotic behavior of the model. It is found
that all the regularization methods except for the dimensional one
actually lead reliable physical predictions for the kaon decay constant,
sigma meson mass and topological susceptibility without restricting
the ultra-violet cutoff below the hadronic scale.
\end{abstract}


\end{frontmatter}


\section{\label{sec:intro}%
Introduction}
Nonet mesons are interesting composite hadronic objects which have been
seriously studied in theoretical and experimental particle physics. The
elementary objects composing mesons are quarks and gluons, and the first
principle theory of them is quantum chromodynamics (QCD). Then one of
our goals is to explain all the information on hadrons from QCD. The most
reliable approach is to consider the discretized version of QCD, called the
lattice QCD, whose technologies are developing day by day. It is, however,
still difficult to study hadrons from the first principle, so the approaches
by using some effective models become one of our options.

In this paper we employ the Nambu--Jona-Lasinio (NJL) model~\cite{NJL}
being one of frequently used models for the investigations of hadronic
particles. The three-flavor model with $U_{\rm A}(1)$
anomaly~\cite{Weinberg:1975ui} called Kobayashi--Maskawa--'t Hooft
(KMT) term~\cite{Kobayashi:1970ji}
successfully describes the nonet meson properties (for reviews, see,
e.g.,~\cite{Vogl:1991qt, Klevansky:1992qe, Hatsuda:1994pi,
Rehberg:1995kh, Buballa:2005rept, Huang:2004ik}).
The model is not renormalizable, since it contains the higher dimensional
operators, four- and six-point fermion interactions. Therefore the model
predictions inescapably depend on the regularization procedures. Also, the
model shows parameter dependence in each regularization method.
Then we have launched a plan to perform the systematical analyses on
both the regularization and parameter dependence.

Here we are going to study the model with five regularizations: the
three-dimensional (3D) and four-dimensional (4D) sharp cutoff schemes,
Pauli-Villars (PV), the proper-time (PT) and the dimensional regularizations
(DR), as the straightforward extension of the work with the two-flavor
model~\cite{Inagaki:2015lma}. The 3D cutoff drops the higher momentum
contribution in the space direction, which is the most frequently used method
due to its simple physical interpretation and nice numerical behavior.
Similarly, the 4D cutoff method kills the amplitudes from higher momentum
in the four-dimensional Euclidean momentum space. The PV way reduces
high momentum contribution by subtracting the amplitudes from virtual heavy
particles~\cite{Pauli:1949zm, Itzykson1980, Cheng1984}. The PT method
introduces the exponentially dumping factor in the integral, then make
divergent loop integrals finite~\cite{Itzykson1980, Schwinger:1951nm}.
The DR prescription modifies an integral kernel through changing the
space-time dimension so as to make divergent integrals finite.
The model has been examined
in detail with various regularizations, see, e.g., for the
4D~\cite{Klevansky:1992qe, Hatsuda:1994pi, Krewald:1991tz, jafarov:2004,
jafarov:2006},
PV~\cite{ Klevansky:1992qe, Jaminon:1989ix, Kahana:1992jm, Osipov:2004bj,
Moreira:2010bx},
PT~\cite{Suganuma:1990nn, Klimenko:1991he, Gusynin:1994re,
Inagaki:1997nv, Inagaki:2004ih, Inagaki:2003yi, Inagaki:2003ac, Cui:2014hya}
and
DR~\cite{Krewald:1991tz, jafarov:2004, jafarov:2006, inagaki:2008,
Inagaki:2010nb, Inagaki:2011uj, Inagaki:2012re, Inagaki:2013hya, Inagaki:2014kta}.
The NJL model is regarded as a low energy effective model of QCD; 
it is the simplest model to induce dynamical chiral symmetry breaking and often
applied to investigate physics near the QCD phase transition. To apply the model
to the nonet mesons $\eta$ and $\eta'$ mesons may be not light enough compared
with the QCD scale. Since the model loses validity at higher energy, it should
be essential to evaluate the safety and effectiveness of the model with the
regularization procedures. It is to be noted that the model has non-negligible
parameter dependence even within the same regularization
procedure~\cite{Inagaki:2015lma}. In particular, some physical
quantities, such as the transition temperature on the chiral symmetry breaking,
are crucially affected by the model parameters. Moreover, there exists some
room for the choice of parameters since input physical quantities for setting
the parameters are usually less than the number of the parameters, then
several parameter sets are employed depending on working
groups~\cite{Hatsuda:1994pi, Rehberg:1995kh, Krewald:1991tz, Osipov:2004bj,
Cui:2014hya}. Therefore it is also important to test the parameter dependence on
the model predictions. A lot of works have been devoted to the searches on
the model parameters with various regularizations. For the sake of  seeing
the regularization and parameter dependence on the physical quantities, 
in this article, we shall perform the systematical parameter fitting in the three
flavor model.

The paper is constructed as follows: Section~\ref{sec:model} presents the model
treatments and the regularization procedures. We will carry on the detailed
parameter fitting in Sec.~\ref{sec:fit}. Section~\ref{sec:prediction} is devoted to the
investigations on the physical predictions. We give some discussions on the
parameter fitting in Sec.~\ref{sec:discussion}. Some concluding remarks
are put in Sec.~\ref{sec:conclusion}. Appendix shows the explicit equations for
the meson properties and topological susceptibility.

\section{\label{sec:model}%
Model and regularizations}
We start from the model Lagrangian then derive the effective potential 
and the gap equations in the leading order of the $1/N_c$ expansion.
Since the integrals appearing in the gap equations involve divergent
contributions, regularization procedures should be introduced to define
the finite integral. Here we consider the three- and four-dimensional
momentum cutoff schemes, Pauli-Villars, the proper-time and the 
dimensional regularizations. The explicit forms of the gap equations
are shown in Sec.~\ref{subsec:model} and the formula for several 
regularization methods are derived in Sec.~\ref{subsec:regularization}.

\subsection{\label{subsec:model}%
The NJL model}
The Lagrangian of the three-flavor NJL model is given by
\begin{eqnarray}
  \mathcal{L}_{\mathrm{NJL}}
   = \overline{q} \left( i\gamma_{\mu}\partial^{\mu}
       - \hat{m}\right) q 
       + G \sum_{a=0}^{8}
           \left[
             \bigl(  \overline{q} \lambda_a q \bigr)^2
           + \bigl( \overline{q} \,i \gamma_5 \lambda_a q \bigr)^2
          \right]
       -K \left[ \det\overline{q}_i (1-\gamma_5) q_j 
     +{\rm h.c.\ } \right] ,
  \label{NJL}
\end{eqnarray}
where $q$ represents quark fields for up, down, and strange, $\hat{m}$
indicates the diagonal mass matrix for the current quarks
$\hat{m}=\{m_u, m_d, m_s \}$, $G$ and $K$ are the four- and six-point
couplings, $\lambda_a$ are the Gell-Mann matrices with
$\lambda_0 = \sqrt{2/3}\, \mbox{1}\hspace{-0.25em}\mbox{l}$   
in the flavor space, and the determinant is taken in the flavor space
leading so-called  Kobayashi--Maskawa--'t Hooft (KMT)
term~\cite{Kobayashi:1970ji}. In QCD the $U_{\rm A}(1)$ symmetry is
broken by the anomaly.
The KMT term explicitly breaks the $U_{\rm A}(1)$ symmetry, and
plays dominant role on the mixture between light and strange quarks 
which will be discussed in detail with the actual numerical analyses.

The mean-field approximation,
$\langle \bar{q_i} q_i \rangle \simeq \phi_i$,
helps us to have the following linearized Lagrangian,
\begin{eqnarray}
 && \hat{\mathcal{L}}
   = \bar{q} (i\gamma_{\mu}\partial^{\mu} -\hat{m}^{*}) q
     -2G(\phi_u^2 +\phi_d^2 +\phi_s^2) + 4K \phi_u \phi_d \phi_s,
	 \label{l:aux}
\end{eqnarray}
where $\hat{m}_i^{*}$ indicates the diagonal matrix whose
elements are the constituent quark masses
\begin{eqnarray}
 m^*_u &=& m_u -4G \phi_u +2K \phi_d \phi_s,  \label{eq:m_u}\\
 m^*_d &=& m_d -4G \phi_d +2K \phi_s \phi_u , \label{eq:m_d}\\
 m^*_s &=& m_s -4G \phi_s +2K \phi_u \phi_d. \label{eq:m_s}
\end{eqnarray}
One can obtain the effective potential, $\Omega=-\ln Z/V$,
where $V$ represents the volume of the system, and $Z$ is the
partition function,
\begin{eqnarray}
  Z = \int \mathcal{D} [q]
        \exp \left[  i \int {\mathrm d}^4 x  \tilde{\mathcal{L}} 
              \right].
\end{eqnarray}
The explicit form of the effective potential becomes
\begin{eqnarray}
  \Omega &=& \Omega_{\phi} + \Omega_{q} , \label{eq:omega0}\\
  \Omega_{\phi}
   &=& 2G(\phi_u^2 + \phi_d^2 + \phi_s^2) 
          -4K \phi_u \phi_d \phi_s, \\
  \Omega_{q} 
   &=&  - \sum_i {\rm tr} \int \frac{{\mathrm d}^4 q}{i(2\pi)^4} 
            \ln \left( q_\mu \gamma^\mu -m_i^* + i\varepsilon \right),
  \label{eq_Omega^0}
\end{eqnarray}
where the trace is taken in the color and spinor indices.
(See, for a review~\cite{Hatsuda:1994pi}.)

The gap equations are derived by differentiating the thermodynamic
potential by the order parameter, $\phi_i$,
\begin{equation}
  \frac{\partial \Omega}{\partial \phi_i} = 0,
  \label{eq:gap0}
\end{equation}
whose solutions give the extremum points of the potential. Note that
one should be careful when the equations have several extremum
points, in which case the direct search of the global minimum by evaluating
the potential itself is necessary. Substituting Eq.~(\ref{l:aux}) with
Eq.~(\ref{eq:omega0}) into Eq.~(\ref{eq:gap0}), we obtain
\begin{equation}
   \phi_i
  = -i{\rm tr} S_i 
  =   {\rm tr} \int \frac{\md^4 q}{i(2\pi)^4}
  \frac{1}{  q_\mu \gamma^\mu - m^*_i + i\varepsilon} ,
\label{eq:gap}
\end{equation}
where $S_i$ represents the propagator for the constituent quarks. As obviously
seen from the above form, the expression for $\phi_i$ quadratically
diverges, so the regularization is needed for the sake of obtaining finite physical
quantities. The concrete procedures of the regularizations will be discussed
in the next subsection.

\subsection{\label{subsec:regularization}%
Regularization procedures}
Having presented the model treatment for analyzing the chiral condensate, we
may now be ready for presenting on the regularization prescription used to
obtain the finite physical predictions. As mentioned in the introduction,
we shall be studying five regularization procedures: the 3D cutoff, 4D cutoff,
Pauli-Villars, proper-time and dimensional regularizations.

In our present investigations, there are two types of divergent integrals to be
made finite by some regularizations. The problematic integrals are
\begin{eqnarray}
  && i{\rm tr} S_i 
  = -{\rm tr} \int \frac{\md^4 q}{i(2\pi)^4}
  \frac{1}{  q_\mu \gamma^\mu - m^*_i + i\varepsilon}, \\
  && I_{ij}(p^2)= {\rm tr} \int  \frac{\md^4 q}{i(2\pi)^4}
  \frac{1}{ \bigl[ (q+p/2)^2-m_i^{*2} + i\varepsilon \bigr] \bigl[ (q-p/2)^2-m_j^{*2}
  + i\varepsilon \bigr]},
\end{eqnarray}
where $I_{ij}(p^2)$ appears when one evaluates the meson properties; the
derivations of the meson properties are presented in \ref{app:meson}.
These divergent integrals should become finite by the regularizations.

\subsubsection{\label{subsubsec:3D}%
Three dimensional-momentum cutoff}
The three dimensional-momentum cutoff is the way to introduce the momentum
cutoff in the three dimensional space momentum direction shown as
\begin{equation}
  \int \frac{\md^4 q}{(2\pi)^4}
  \to 
  \int \frac{\md q_0}{2\pi}
  \int^{\Ltd}_0 \frac{q^2 \md q}{(2\pi)^3}\int \md \Omega_{3}.
\end{equation}
This is the most frequently used method in the NJL analyses due to the
straightforward physical interpretation and its convenience for numerical
calculations.

By introducing the cutoff scale, $\Ltd$, we have the following simple expressions
for $i{\rm tr}S$ and the quark loop amplitude,
\begin{eqnarray}
  && i{\rm tr}S^{\rm 3D}_i 
  =  -\frac{N_c m_i^*}{2\pi^2}
    \left( \Ltd \sqrt{\Ltd^2 + m_i^{*2}}
    -m_i^{*2} \ln \frac{\Ltd + \sqrt{\Ltd^2 + m_i^{*2}}}{m_i^*} 
   \right), 
   \label{eq:tr3D}  \\
  && I_{ij}^{\rm 3D} =
  4N_c \int^{\Ltd} \frac{\md^3 q}{(2\pi)^3} 
  \frac{1}{2D_{ij}^{+}}
  \left(  \frac{1}{E_i} +\frac{1}{E_j}
  \right),
\end{eqnarray}
with $D^{+}_{ij} =(E_i + E_j)^2-p^2$ and $E_i = \sqrt{q^2 + m_i^{*2}}$.
These quantities are to be used for the gap equations and the calculations for
the meson properties and the topological susceptibility.

\subsubsection{\label{subsubsec:4D}%
Four dimensional-momentum cutoff}
There is alternative prescription by employing the sharp momentum cutoff;
that is the four dimensional-momentum cutoff method. One introduces
the covariant cutoff scale, $\Lfd$, after going to the Euclidean space
by the Wick rotation,
\begin{equation}
  \int \frac{\md^4 q_{\rm E}}{(2\pi)^4}
  \to 
  \int^{\Lfd}_0 \frac{q_{\rm E}^3 \md q_{\rm E}}{(2\pi)^4}\int \md \Omega_4.
\end{equation}

As in the 3D case, the integrals for $i{\rm tr}S$ can be evaluated analytically
and reads
\begin{eqnarray}
 i {\rm tr} S^{\rm 4D}_i
  = -\frac{N_c m_i^*}{4\pi^2}
  \left[ \Lfd^2  -m_i^{*2} \ln \left( \frac{\Lfd^2+m_i^{*2}}{m_i^{*2}} \right) 
  \right].
  \label{eq:tr4D}
\end{eqnarray}
There arises a complexity for the quark loop integral depending on the
value of $p^2$, then we separate the integral into three
terms,
\begin{eqnarray}
  &&
  I_{ij}^{\rm 4D}(p^2)
  =    I_{ij}^{\rm 4D(1)}(p^2) 
     + I_{ij}^{\rm 4D(2)}(p^2) +I_{ij}^{\rm 4D(3)}(p^2), \label{eq:Iij}\\
  &&
  I_{ij}^{\rm 4D(1)}(p^2)
  = \frac{N_c}{4\pi^2} \left[
      \int_0^1 \md x \, \ln (\Lfd^2 + \Delta_{ij})
      \right], \\
  &&
  I_{ij}^{\rm 4D(2)}(p^2)
  = -\frac{N_c}{4\pi^2} \left[
      \int_0^1 \md x \, \ln (|\Delta_{ij}|)
      \right], \\
  &&
  I_{ij}^{\rm 4D(3)}(p^2)
  = -\frac{N_c}{4\pi^2} \Lfd^2 \left[
      \int_0^1 \md x \, \frac{1}{\Lfd^2 + \Delta_{ij}}
      \right],
\end{eqnarray}
where
\begin{eqnarray}
  && \Delta_{ij} = p^2 [(x-A_{ij})^2 + B_{ij}], \\
  && A_{ij} = \frac{1}{2}\left( 1 + \frac{m_j^{*2}-m_i^{*2}}{p^2} \right), \\
  && B_{ij} = -\frac{1}{4} + \frac{m_i^{*2} + m_j^{*2}}{2p^2}
        - \frac{(m_j^{*2}-m_i^{*2})^2}{4p^4}.
\end{eqnarray}
The integration in the first and third terms,
$I_{ij}^{\rm 4D(1)}$ and $I_{ij}^{\rm 4D(3)}$, in Eq.~(\ref{eq:Iij})
are straightforward since $\Lfd^2 + \Delta_{ij}$ is always positive,
\begin{eqnarray}
  I_{ij}^{\rm 4D(1)}(p^2)
  & = &
     \frac{N_c}{4\pi^2} 
     \biggl[ 
        \ln p^2 -2 + (1-A_{ij}) \ln[(1-A_{ij})^2 + c_{ij}^2] 
        +A_{ij} \ln[ A_{ij}^2 + c_{ij}^2] \nonumber \\
  &&
                 +2c_{ij} \arctan \left( \frac{1-A_{ij}}{c_{ij}} \right)
                 +2c_{ij} \arctan \left( \frac{A_{ij}}{c_{ij}} \right)
     \biggr], \\
  I_{us}^{\rm 4D(3)}(p^2)
  & = & -\frac{N_c}{4\pi^2} \, \frac{\Lfd^2}{c_{ij}p^2}
     \biggl[     \arctan \left( \frac{1-A_{ij}}{c_{ij}} \right)
                 + \arctan \left( \frac{A_{ij}}{c_{ij}} \right)
     \biggr],
\end{eqnarray}
with
\begin{equation}
  c_{ij} = \sqrt{ \frac{\Lfd^2}{p^2} + B_{ij}}. 
\end{equation}
While the second term, $I_{ij}^{\rm 4D(2)}$, needs careful evaluation
since $\Delta_{ij}$ can be negative if $B_{ij}$ becomes negative.
We have for $B_{ij} > 0$, 
\begin{eqnarray}
  I_{ij}^{\rm 4D(2)}(p^2)
  & = & -\frac{N_c}{4\pi^2} 
     \biggl[ 
        \ln p^2 -2 + (1-A_{ij}) \ln[(1-A_{ij})^2 + b_{ij}^2] 
        +A_{ij} \ln[ A_{ij}^2 + b_{ij}^2] \nonumber \\
  && 
                 +2b_{ij} \arctan \left( \frac{1-A_{ij}}{b_{ij}} \right)
                 +2b_{ij} \arctan \left( \frac{A_{ij}}{b_{ij}} \right)
     \biggr],
\end{eqnarray}
with $b_{ij} \equiv \sqrt{|B_{ij}|}$, and for $B_{ij} < 0$,
\begin{eqnarray}
  I_{ij}^{\rm 4D(2)}(p^2) 
  = -\frac{N_c}{4\pi^2} 
      \left[ \ln p^2 -2 + a_{ij}^- \ln (a_{ij}^-) + (1-a_{ij}^-) \ln (1-a_{ij}^-)
               + a_{ij}^+ \ln (a_{ij}^+) + (1-a_{ij}^+) \ln (1-a_{ij}^+) \right] ,
\end{eqnarray}
with $a_{ij}^{\pm} = A_{ij} \pm b_{ij}$.

\subsubsection{Pauli-Villars regularization}
In the Pauli-Villars regularization, one suppresses the divergent integrals
through introducing the frictional force by
\begin{equation}
  \frac{1}{q^2-m^2} \, \longrightarrow \,
  \frac{1}{q^2-m^2}
 -\sum_k \frac{a_k}{q^2-\Lambda_k^2},
\end{equation}
where the cutoff scale is determined by the virtual heavy mass,
$\Lambda_k$.  To obtain the finite functions, we subtract the
integral of $i{\rm tr}S$ and $I$ with the sum of $k=1,2$.
This cutoff scale $\Lambda_k$ is replaced by the common
model cutoff $\Lpv$ after some algebras which makes all the contributions
finite~\cite{Inagaki:2015lma}.

The integration in $i{\rm tr}S$ can be performed analytically by using \
both the 3D and 4D expressions shown in Eqs. (\ref{eq:tr3D}) and
(\ref{eq:tr4D}),
\begin{eqnarray}
 i {\rm tr}S^{\rm PV}_i
  = -\frac{N_c m_i^*}{4\pi^2} 
     \left( \Lpv^2  - m_i^{*2} + m_i^{*2} \ln \frac{m_i^{*2}}{\Lpv^{2}}
     \right),
\label{potential_PV}
\end{eqnarray}
It may be worth showing both the ways for the integral in $I_{ij}$;
one sees in the 3D case,
\begin{eqnarray}
  I_{ij}^{\rm PV(3D)}(p^2)
  &=&  I_{ij}^{\rm 3D}(p^2)
         - \frac{1}{2} I_{i\Lambda}^{\rm 3D}(p^2) 
         - \frac{1}{2} I_{j\Lambda}^{\rm 3D}(p^2)  
         \nonumber \\
  &=&  4N_c \int^{\infty} \!\!\!\! \frac{\md^3 q}{(2\pi)^3} 
      \left[
        \frac{1}{2D_{ij}^{+}}
        \left(
           \frac{1}{E_i} + \frac{1}{E_j}
        \right)
     - \sum_{k=i,j}
        \left\{
        \frac{1}{4D_{k \Lambda}^{+}}
        \left(
           \frac{1}{E_k} + \frac{1}{E_\Lambda}
        \right)
        \right\}
     \right].
\end{eqnarray}
where in $I_{k \Lambda}^{3D}$, the constituent quark masses
are replaced by the cutoff $\Lpv$ as
\begin{eqnarray}
  D_{k \Lambda}^{+} = (E_k + E_{\Lambda})^2-p^2, \quad
  E_{\Lambda}^2 = q^2 + \Lpv^2.
\end{eqnarray}
One also sees in the 4D case,
\begin{eqnarray}
  I_{ij}^{\rm PV(4D)}(p^2)
  =   I_{ij}^{\rm 4D(2)}(p^2)
     - \frac{1}{2} I_{i \Lambda}^{\rm 4D(2)}(p^2)
     - \frac{1}{2} I_{j \Lambda}^{\rm 4D(2)}(p^2).
\end{eqnarray}
where $m_i^{*}$ and $m_j^*$ are replaced by $\Lpv$ as well in
$I_{k \Lambda}^{\rm 4D(2)}$,
\begin{eqnarray}
  &&I_{k \Lambda}^{\rm 4D(2)}(p^2)
  = -\frac{N_c}{4\pi^2} 
     \biggl[ 
        \ln p^2 -2 + (1-A_{k \Lambda}) 
        \ln[(1-A_{k \Lambda})^2 + b_{k \Lambda}^2] 
        +A_{k \Lambda} \ln[ A_{k \Lambda}^2 + b_{k \Lambda}^2] \nonumber \\
    && \qquad \qquad \qquad \qquad
                 +2b_{k \Lambda}
                    \arctan \left( \frac{1-A_{k \Lambda}}{b_{k \Lambda}} \right)
                 +2b_{k \Lambda}
                    \arctan \left( \frac{A_{k \Lambda}}{b_{k \Lambda}} \right)
     \biggr],
\end{eqnarray}
with
\begin{eqnarray}
  && A_{k \Lambda} = \frac{1}{2}\left( 1 + \frac{\Lpv^2 - m_k^{*2}}{p^2} \right), \\
  && B_{k \Lambda} = -\frac{1}{4} + \frac{m_k^{*2} + \Lpv^{2}}{2p^2}
        - \frac{(\Lpv^{2} -m_k^{*2})^2}{4p^4}.
\end{eqnarray}
We have numerically confirmed that these two expressions give the equal
results as they should.

\subsubsection{Proper-time regularization}
In the proper-time regularization, the divergent integrals are made finite
by suppressing the high momentum contributions with the insertion of the
exponentially dumping factor through the following manipulation,
\begin{equation}
  \frac{1}{A^n} \to \frac{1}{\Gamma[n]}\int_{1/\Lpt^2}^{\infty} \md \tau\,\, 
  \tau^{n-1}e^{-A \tau}.
\label{eq:proper}
\end{equation}

The integration in $i{\rm tr}S$ is easily performed,
\begin{eqnarray}
  && i {\rm tr}S^{\rm PT}_i
  = -\frac{N_c m_i^*}{4\pi^2}
   \left[ \Lpt^2 e^{-m_i^{*2}/\Lpt^2} 
   + m_i^{*2} Ei(-m_i^{*2}/\Lpt^2)
   \right] .
  \label{gap_pt}
\end{eqnarray}
In the similar manner treated above, $I_{ij}$,
\begin{eqnarray}
  &&I_{ij}^{\rm PT}(p^2)
  = \frac{N_c}{4\pi^2}
    \int_0^1 \!\! \md x  \int_{1/\Lpt^2}^{\infty} \md \tau \,
    \frac{1}{\tau} e^{- \Delta_{ij} \tau},
\end{eqnarray}
should be calculated depending on the sign of $\Delta_{ij}$.  
It becomes for $B_{ij}>0$,
\begin{eqnarray}
  &&I_{ij}^{\rm PT}(p^2)
  = -\frac{N_c}{4\pi^2} \int_0^1 \md x \, Ei(-\Delta_{ij}/\Lpt^2),
\end{eqnarray}
and for $B_{ij} < 0$,
\begin{eqnarray}
  && I_{ij}^{\rm PT}(p^2)
  = I_{ij}^{\rm PT(1)}(p^2) 
      + I_{ij}^{\rm PT(2)}(p^2) 
      + I_{ij}^{\rm PT(3)}(p^2) ,\\  
  && I_{ij}^{\rm PT(1)}(p^2)
  = -\frac{N_c}{4\pi^2} \int_0^{\alpha_{-}} \md x \, 
                     Ei(-\Delta_{ij}/\Lpt^2) , \\
  && I_{ij}^{\rm PT(2)}(p^2)
  = -\frac{N_c}{4\pi^2} \left[ \int_{\alpha_{-}}^{\alpha_{+}} \md x \, 
                     Ei(\Delta_{ij}/\Lpt^2) 
     +i \int_{-\pi/2}^{\pi/2} \md \, 
                     \theta e^{i \Delta_{ij}e^{i\theta}/\Lpt^2 }
    \right], \\
  && I_{ij}^{\rm PT(3)}(p^2)
  = -\frac{N_c}{4\pi^2} \int_{\alpha_{+}}^1 \md x \, 
                     Ei(-\Delta_{ij}/\Lpt^2) ,
\end{eqnarray}
with $a_{ij}^{\pm} = A_{ij} \pm \sqrt{B_{ij}}$ as already defined.

\subsubsection{Dimensional Regularization}
In the dimensional regularization, the integral kernel is modified by
changing the space-time dimensions,
\begin{equation}
  \int \frac{\md^4 q}{(2\pi)^4}
  \to M_0^{4-D}
  \int \frac{\md^D q}{(2\pi)^D},
\end{equation}
with the mass scaled parameter $M_0$ which plays the role to maintain
the mass dimensions of physical quantities. Note that $D$ should be
restricted to $2 < D <4$ so that one gets finite quantities.

The Feynman integral prescription enables us to obtain the
following results,
\begin{eqnarray}
&& i{\rm tr} S^{\rm DR}_i
  = -\frac{N_c  M_0^{4-D} m_i^*} {(2\pi)^{D/2}} 
      \Gamma \left( 1-\frac{D}{2} \right)
      (m_i^{*2})^{D/2-1}, \\
&&I_{ij}^{\rm DR}(p^2)
  = \frac{N_cM_0^{4-D}}{(2\pi)^{D/2}} \Gamma \left(2-\frac{D}{2} \right)
    \int_0^1 \!\! \md x \Delta_{ij}^{D/2-2} .
\end{eqnarray}
In the DR the mass dimensions of $G$ and $K$ are shifted for its consistency.
In order to compare the couplings with the other regularizations we adjust the
mass dimension of the couplings and write $GM_0^{4-D}$ and $KM_0^{2(4-D)}$
as $G$ and $K$, respectively.

We have now aligned all the required integrals for the evaluation of the
meson properties in the current study. We will then perform the parameter
fitting in the next section.

\section{\label{sec:fit}%
Parameter fitting}
The main issue of this paper is to fit the model parameters systematically.
The suitable parameters can be obtained to reproduce nonet meson
properties in each regularization. After explaining fitting conditions, we 
fixed the parameters as changing
$m_u$ in each regularization and examine the $m_u$ dependence on
the fitted parameters. Next, we replot the obtained results as the
functions of $\Lambda$ and consider the cutoff dependence on the
model parameters.

\subsection{\label{subsec:procedure}%
Fitting procedure}
Since the mass difference between $m_u$ and $m_d$ is negligibly
smaller compared with the hadronic scale, we equalize these masses,
$m_d = m_u$ for simplicity. After the equalization, the model has
five and six parameters in the 3D, 4D, PV, PT cases, and the DR as
alined below:
\begin{eqnarray}
  {\rm 3D,\, 4D,\, PV,\, PT}&{\rm :}&
  \Lambda,\,\, m_u,\,\,  m_s,\,\, G,\,\, K,  \nonumber  \\
  {\rm DR}&{\rm :}& M_0,\,\, D,\,\, m_u,\,\,  m_s,\,\, G,\,\, K, \nonumber
\end{eqnarray}
where the scales of the models are determined by the cutoff $\Lambda$
in the 3D, 4D, PV and PT cases. While in the case with the DR, there is no
direct counterpart to the cutoff scale, then we choose the mass scale with
the factor $4\pi$, namely $\Lambda_{\rm DR} \equiv 4\pi M_0$,
to compare with the other regularizations~\cite{Krewald:1991tz}.
We discuss this point in subsection~\ref{subsec:onDR} in more detail.

The parameters should be set to reproduce physical quantities so that the
models effectively describe real hadron physics. We tune the
model parameters with fitting the following observables~\cite{Agashe:2014kda},
\begin{eqnarray}
  m_\pi = 138{\rm MeV}, \quad
  f_{\pi} =92{\rm MeV}, \quad
  m_{\rm K} = 495{\rm MeV}, \quad
  m_{\eta^{\prime}}=958{\rm MeV}, \nonumber
\end{eqnarray}
as the important ingredients from the experimental observations.
There exists one additional parameter in the DR case, so we select one
more observable,
\begin{eqnarray}
  m_{\eta} = 548{\rm MeV}. \nonumber
\end{eqnarray}
Note that the number of observables is still one less than the number of
parameters. We use $m_u$ as a input parameter. Then the number of 
the remaining parameters are four (and five for DR).

The parameter fitting is technically involved, since one has to solve
four equations,
\begin{eqnarray}
 {\mathcal F}_{\pi}  & \equiv & 1-2K_{\pi} \Pi_{\pi}(m_\pi^2)
   = 0, \label{eq0:Fpi}\\
 {\mathcal F}_{f}  & \equiv &
   f_{\pi} - m^*_u  g_{\pi qq}  I_{u}(0) =0, \label{eq0:Ff}\\
 {\mathcal F}_{\rm K}  & \equiv & 1-2K_{\rm K} \Pi_{\rm K}(m_{\rm K}^2)
   = 0, \label{eq0:Fk}\\
 {\mathcal F}_{\eta^{\prime}}  & \equiv &
      \det \left[ 1-2 \hat{\Pi}(m_{\eta^{\prime}}^2) \hat{\bf K}  \right]
   = 0,
\label{eq0:Fetap}
\end{eqnarray}
under the stationary condition where the two gap equations (\ref{eq:gap})
are satisfied. ${\mathcal F}_\pi$, ${\mathcal F}_f$, ${\mathcal F}_{\rm K}$
and ${\mathcal F}_\eta$ give the equations for the
pion mass, the pion decay constant, the kaon mass and the $\eta^{\prime}$
mass, Eqs.~(\ref{eq:meson_mass}), (\ref{eq:pi_decay}) and (\ref{eq:eta}).
We need one more condition in the DR method as mentioned above,
there we use the equation
\begin{eqnarray}
  {\mathcal F}_{\eta} \equiv \det [ 1-2 \hat{\Pi}(m_{\eta}^2) \hat{\bf K} ] = 0,
  \label{eq0:Feta}
\end{eqnarray}
for $\eta$ meson mass.

As the main task of this paper, we shall systematically solve these 
Eqs.~(\ref{eq0:Fpi}), (\ref{eq0:Ff}), (\ref{eq0:Fk}), (\ref{eq0:Fetap}),
(and (\ref{eq0:Feta})) with two gap equations then obtain the suitable
parameters for each model. It maybe worth mentioning that, although
seven equations are introduced in the DR case with additional mass equation
for $\eta$, the issue is also reduced to the problem with six-coupled equations
since the condition for determining the mass scale $M_0$ can be treated
separately~\cite{Inagaki:2010nb}.

\subsection{\label{subsec:para_mu}%
Fitted parameters with respect to $m_u$}
Since when we perform the parameter fitting, we first fix $m_u$ as free
parameter then evaluate the remaining parameters in our numerical code,
we observe the behavior of the other model parameters
as functions of $m_u$ here.

\begin{figure}[h!]
\begin{center}
   \includegraphics[width=7.5cm,keepaspectratio]{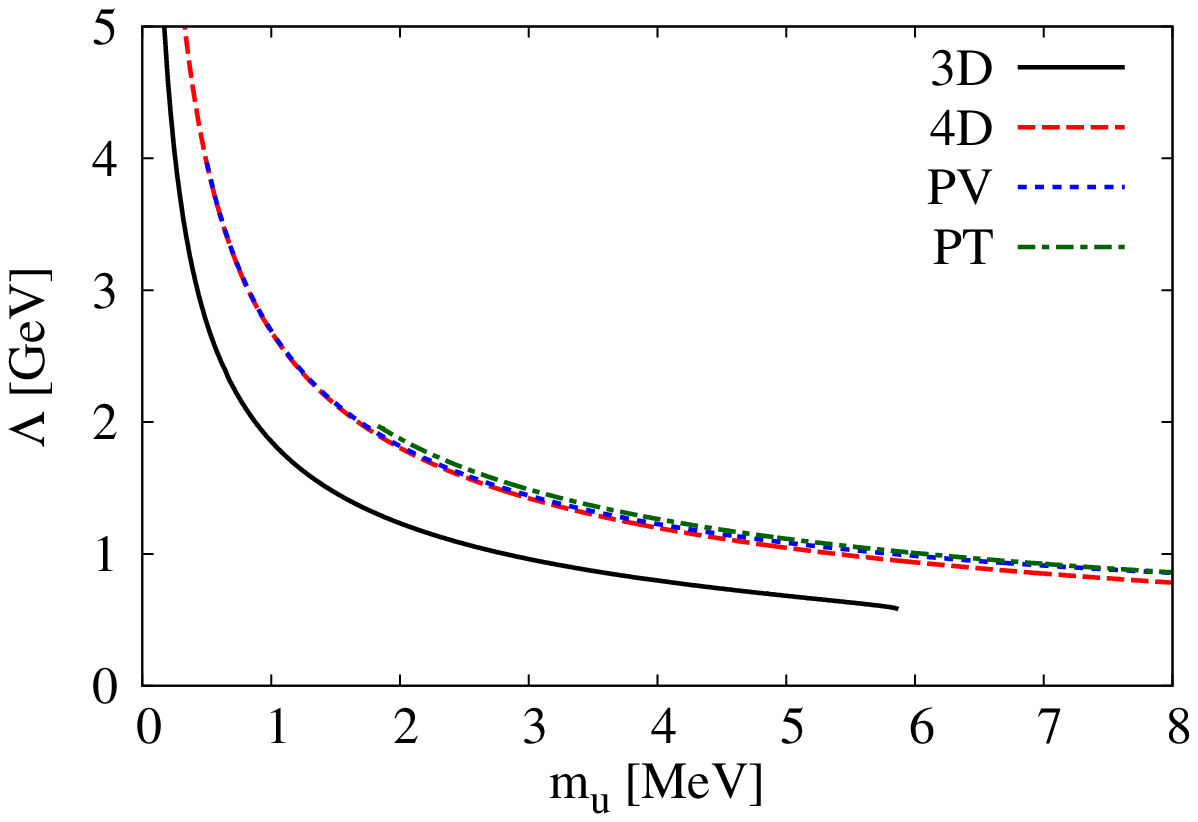}
   \includegraphics[width=7.5cm,keepaspectratio]{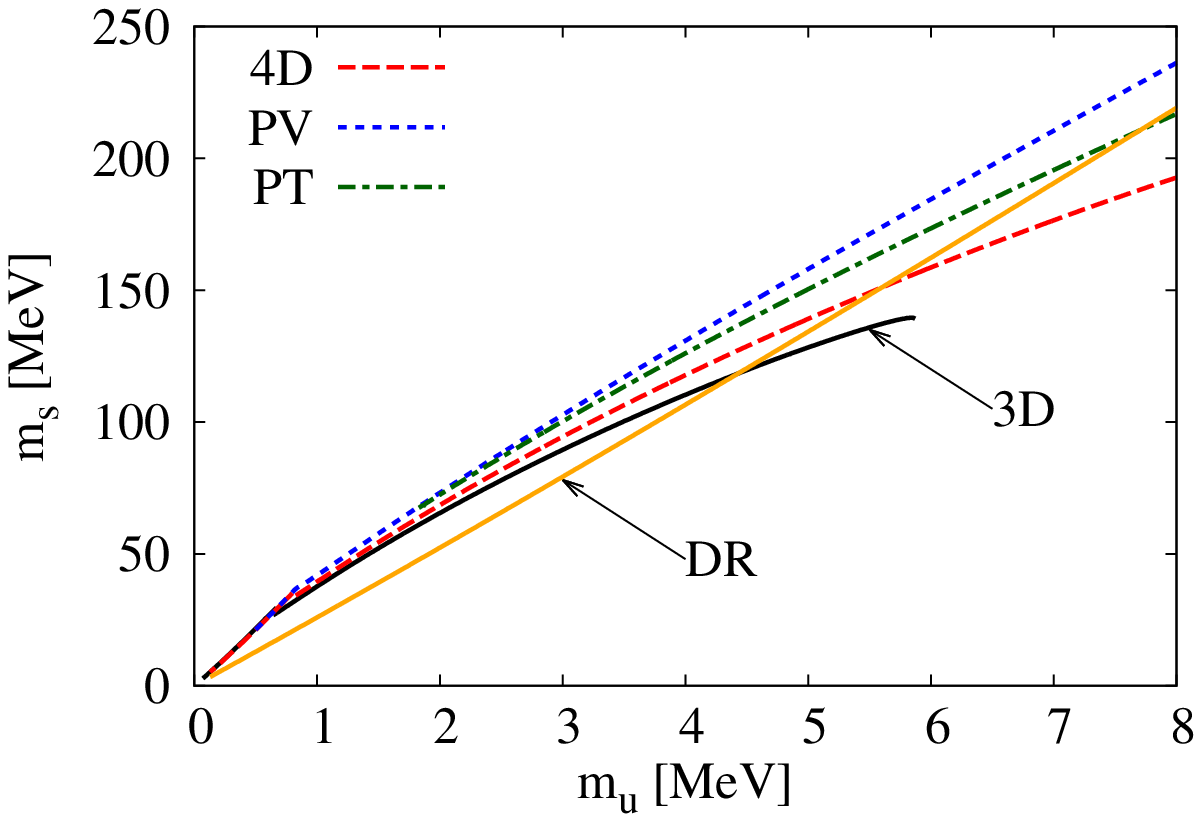}
   \includegraphics[width=7.5cm,keepaspectratio]{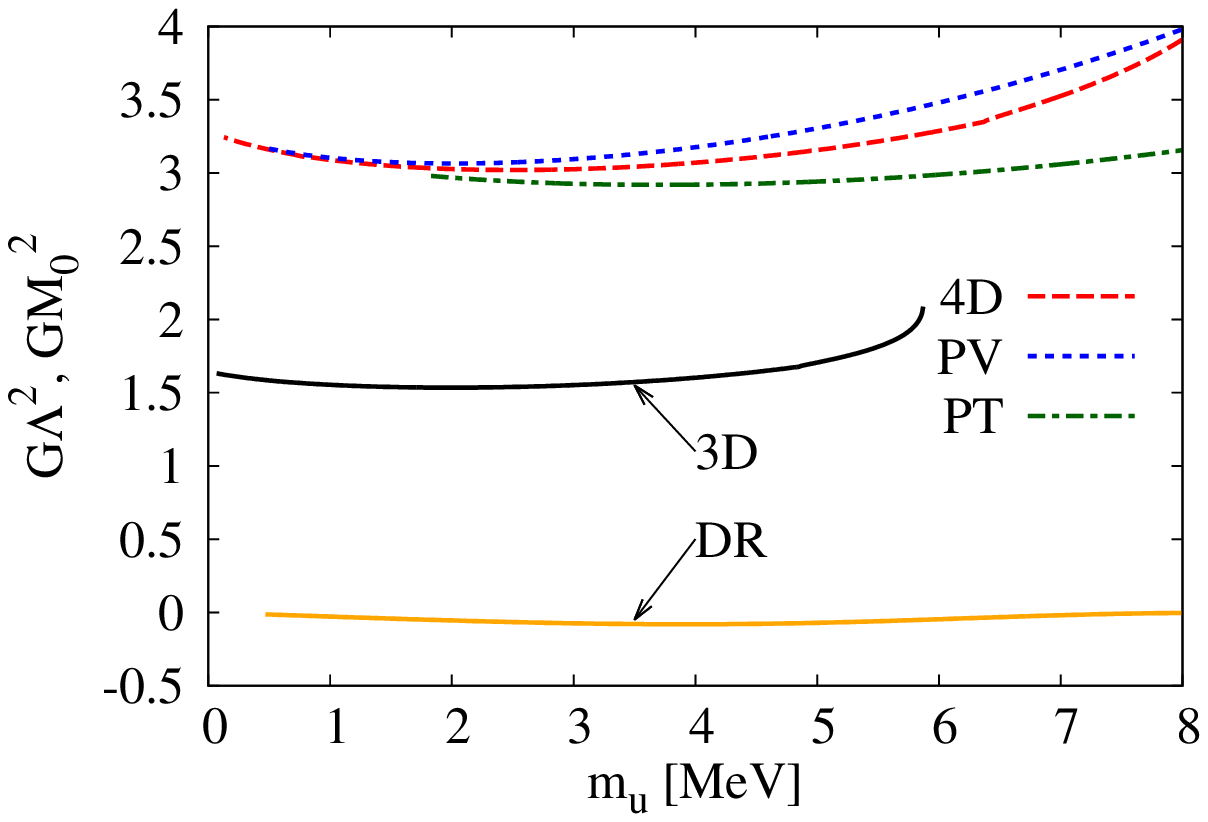}
   \includegraphics[width=7.5cm,keepaspectratio]{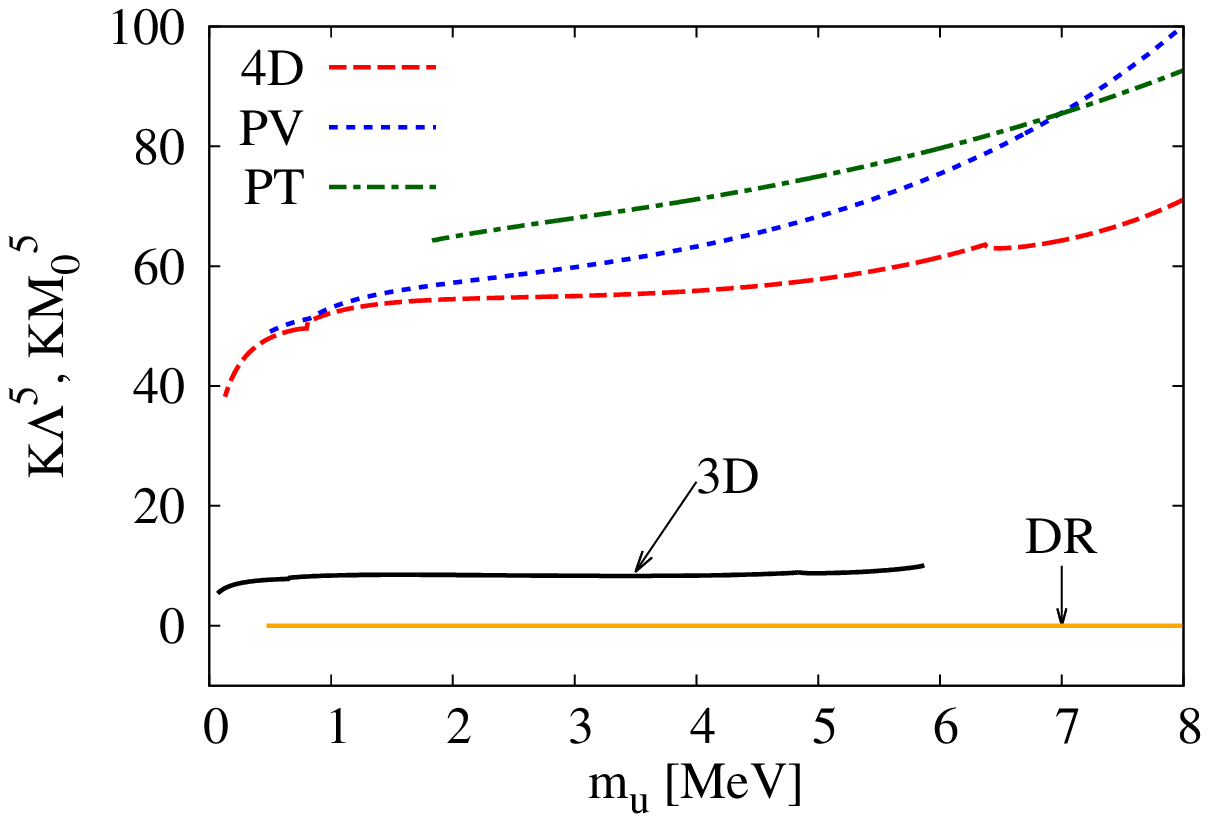}
   \includegraphics[width=7.5cm,keepaspectratio]{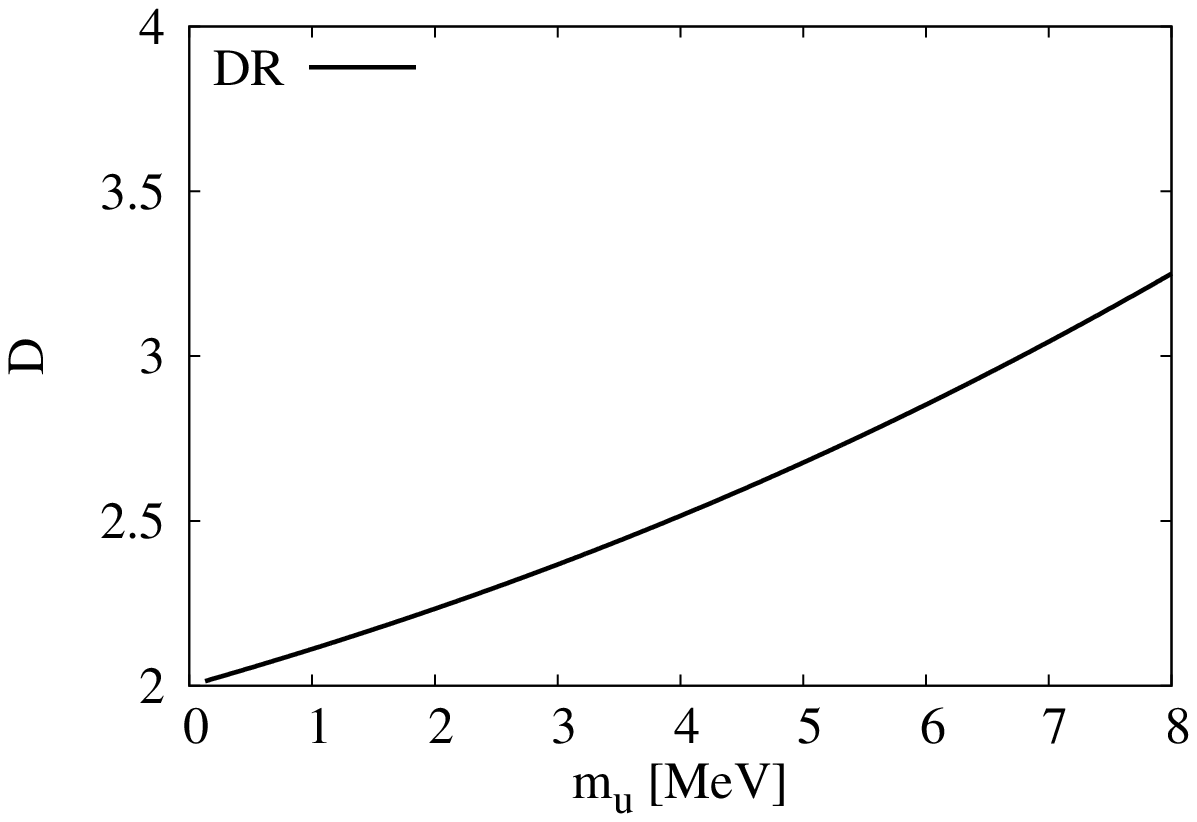}
    \includegraphics[width=7.5cm,keepaspectratio]{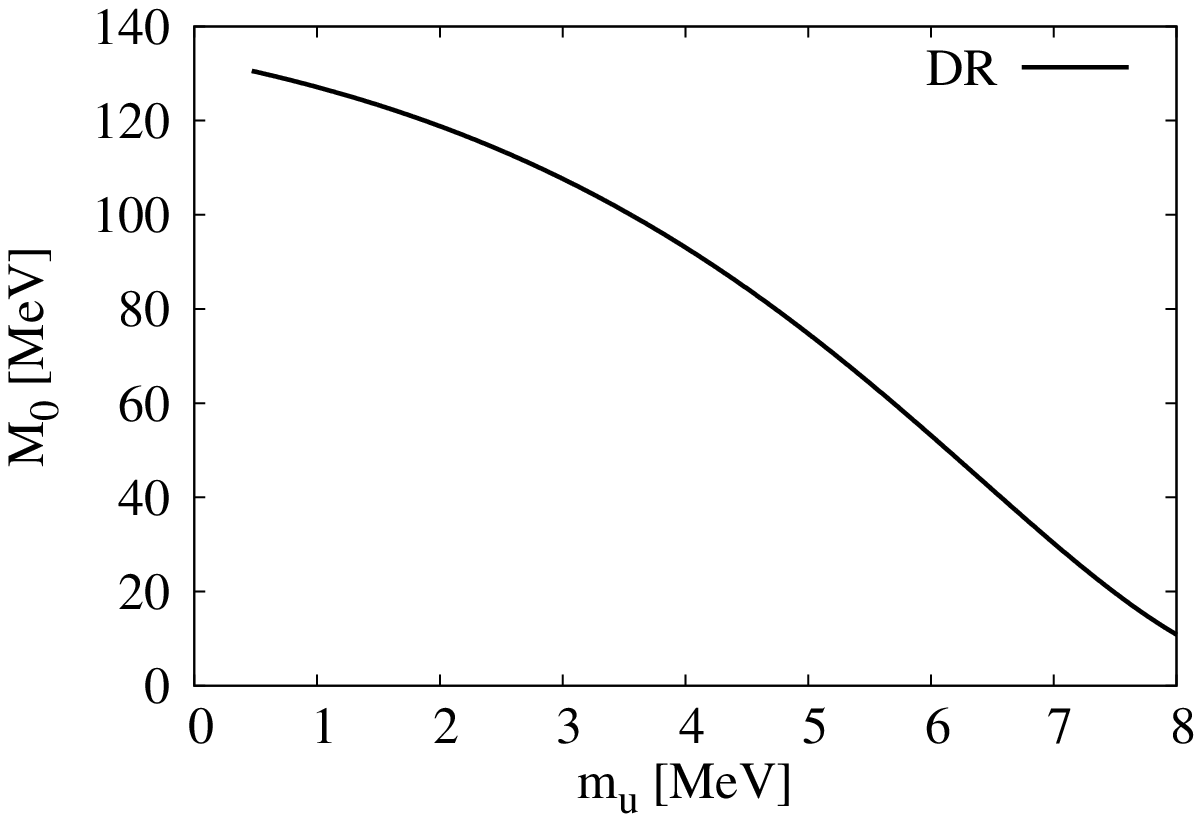}
   \caption{\label{fig:para_mu}
     Parameters as the functions of $m_u$.
     }
\end{center}
\end{figure}
Figure \ref{fig:para_mu} shows the resulting parameters
with respect to $m_u$.
One observes that $\Lambda$ monotonically
decreases with increasing $m_u$, while $m_s$ increases according
to $m_u$, from upper left and right panels. There are no set of solutions
under the conditions discussed in Sec. \ref{subsec:procedure} in the
3D, 4D, PT and DR for large (small) $m_u$ region. Note that there are two
sets of solutions in the DR case, and we draw the curves for the higher
dimension case here.

Middle two panels display the results of the coupling strengths, $G$ and $K$,
in which we note that these quantities stay almost constant for each regularization
in $0\lesssim m_u \lesssim 4$MeV, and the curves in the 4D, PV and PT are
rather close comparing to the other methods in this region.  The DR case
shows peculiar feature; $G$ is negative for all $m_u$ and $K$ is considerably
smaller than the other four regularizations. We can say these are characteristic
aspects within the DR case. The DR has two parameters
on behalf of the cutoff scale, $D$ and $M_0$, where
we observe $D$ increases and the mass scale parameter $M_0$ decreases
with respect to $m_u$, respectively. We will present more detailed discussions
on the behaviors of the obtained parameters in the next subsection with
choosing the cutoff scale $\Lambda$ as the horizontal axis .

\subsection{\label{subsec:para_L}%
Fitted parameters with respect to $\Lambda$}
It is physically more appealing that we redisplay the parameters
as the function of the model scale, $\Lambda$, with obtained parameters,
since one can observe  how the physical quantities effectively flow
according to the model scale in the current  effective field theory.

\begin{figure}[h!]
\begin{center}
   \includegraphics[width=7.5cm,keepaspectratio]{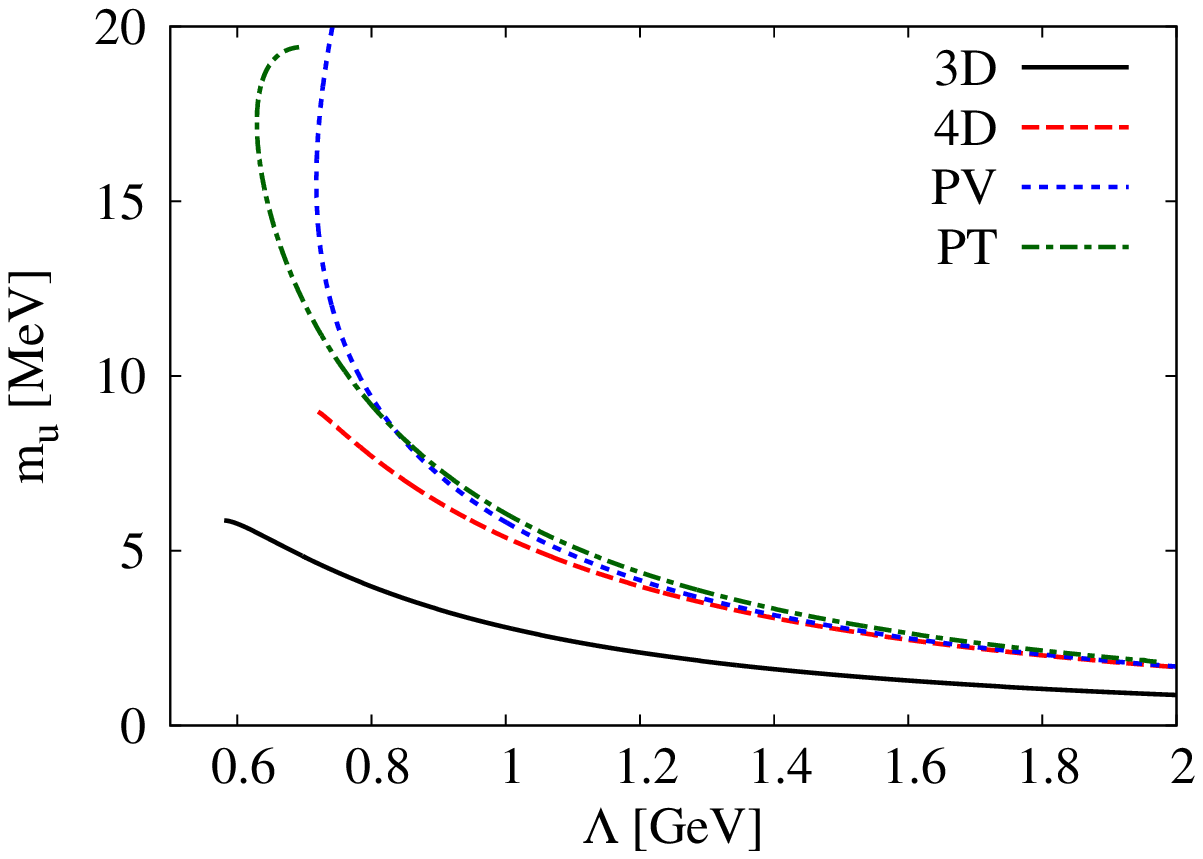}
   \includegraphics[width=7.5cm,keepaspectratio]{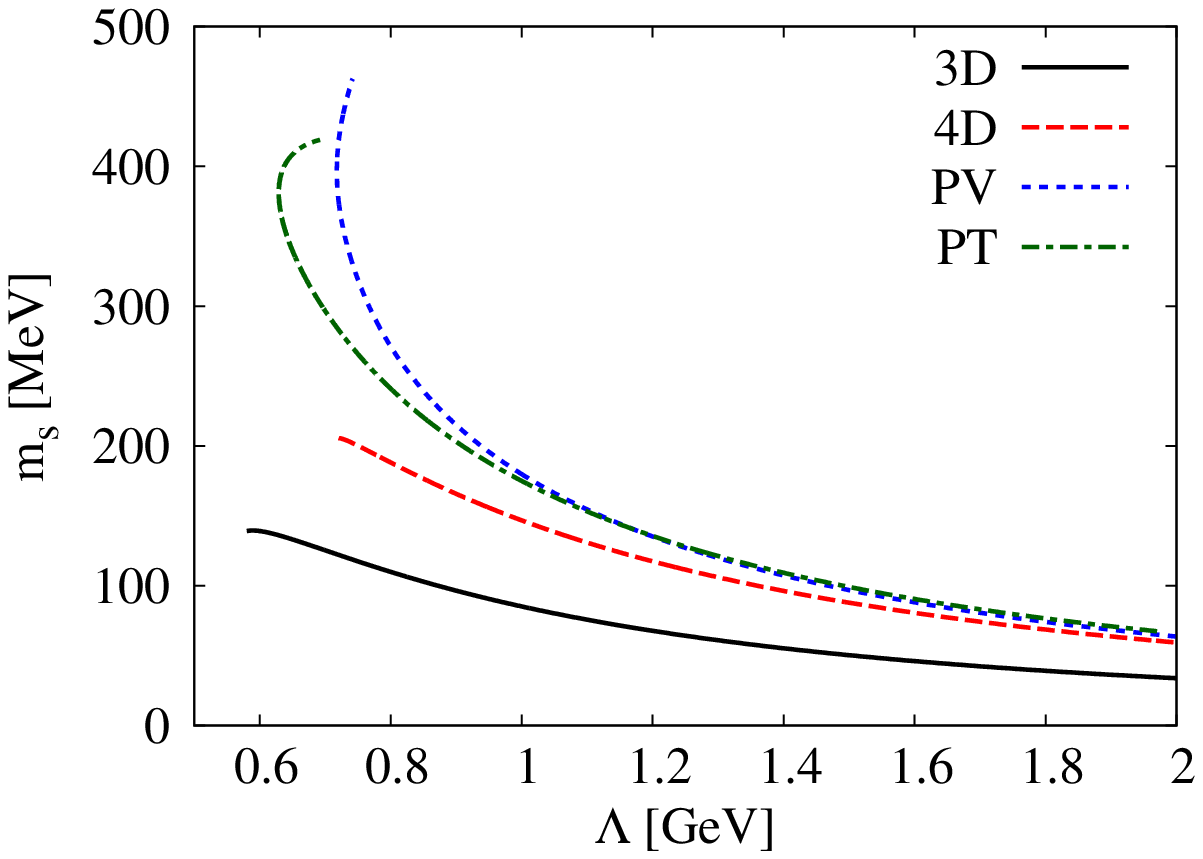}
   \includegraphics[width=7.5cm,keepaspectratio]{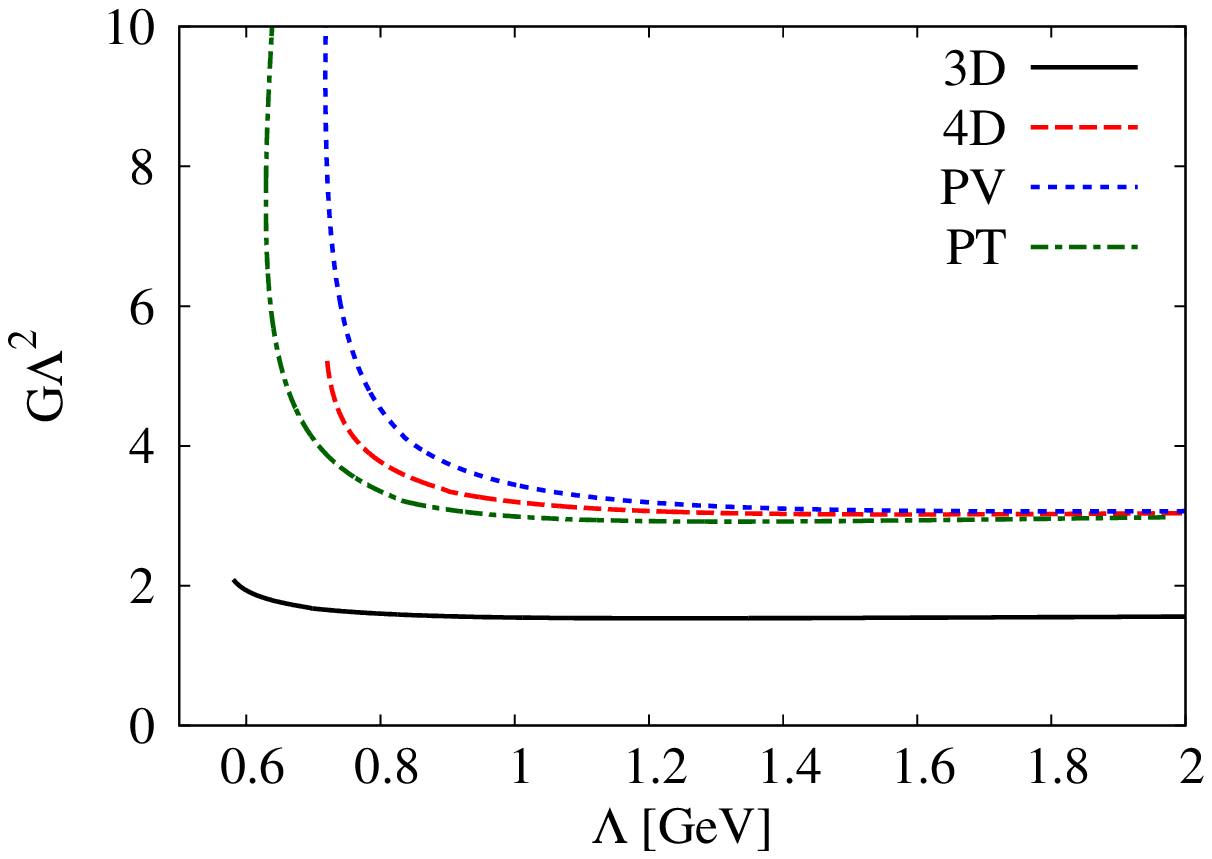}
   \includegraphics[width=7.5cm,keepaspectratio]{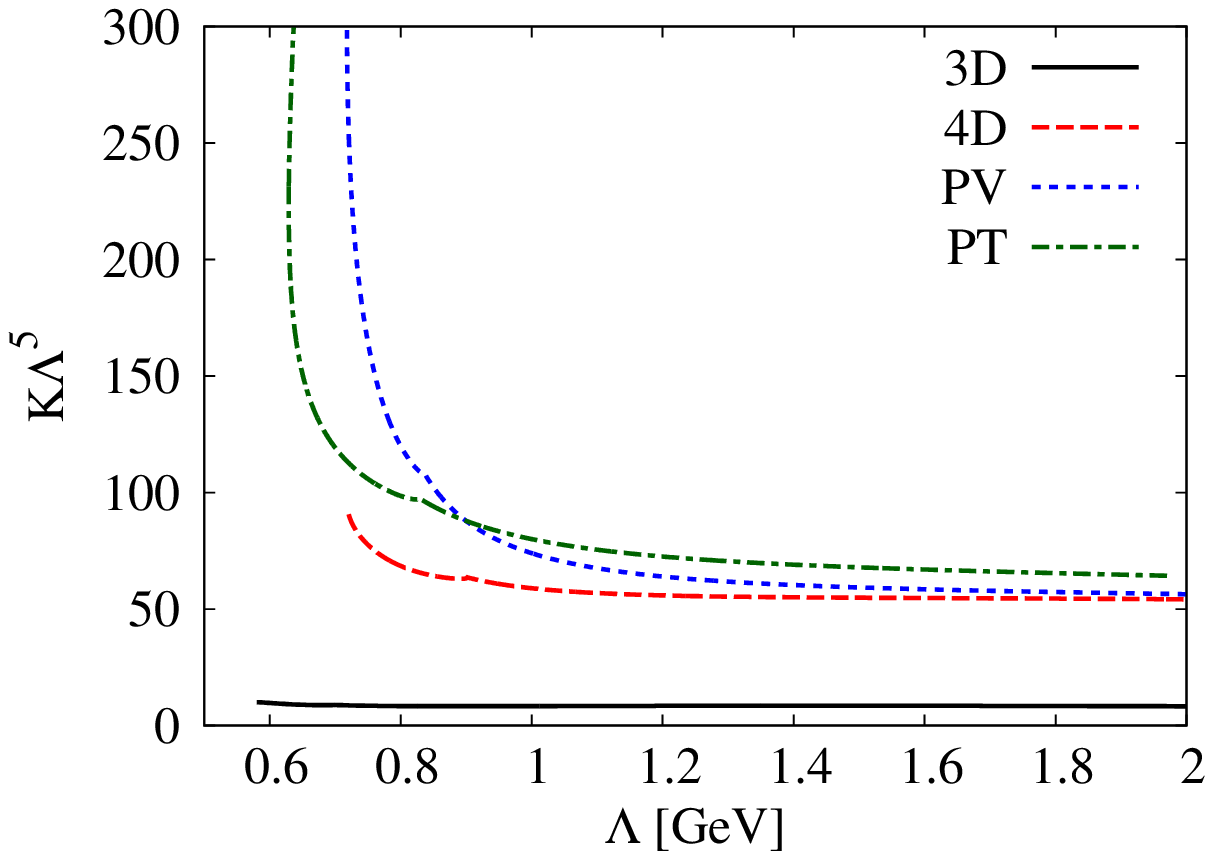}
   \caption{\label{fig:para}
     Parameters in the 3D, 4D, PV, PT.}
\end{center}
\end{figure}
Figure \ref{fig:para} draws four parameters as the functions of the cutoff
scale $\Lambda$ in the 3D, 4D, PV and PT. One sees that $m_u$ and $m_s$
decrease when $\Lambda$ becomes larger, while two couplings $G$ and
$K$ decrease for small $\Lambda$ then approach almost constant for
large $\Lambda$. The PV and PT cases show the multi-valued functions for small
cutoff scale, which is the characteristic feature seen in the PT case being
also observed in the preceding analyses \cite{Inagaki:2003yi}. 
Qualitatively, we can say that all the regularization procedures show
similar curves where each parameter becomes smaller when one takes larger
cutoff scale. On the other hand quantitatively, the three regularizations, 4D, PV and
PT lead close curves, whose values are much larger than the case in the 3D.
This can easily be understood, because the 4D, PV and PT methods have
the mathematically similar structure, while the 3D cutoff separately perform
the time and space integrals.

\begin{figure}[h!]
\begin{center}
   \includegraphics[width=7.5cm,keepaspectratio]{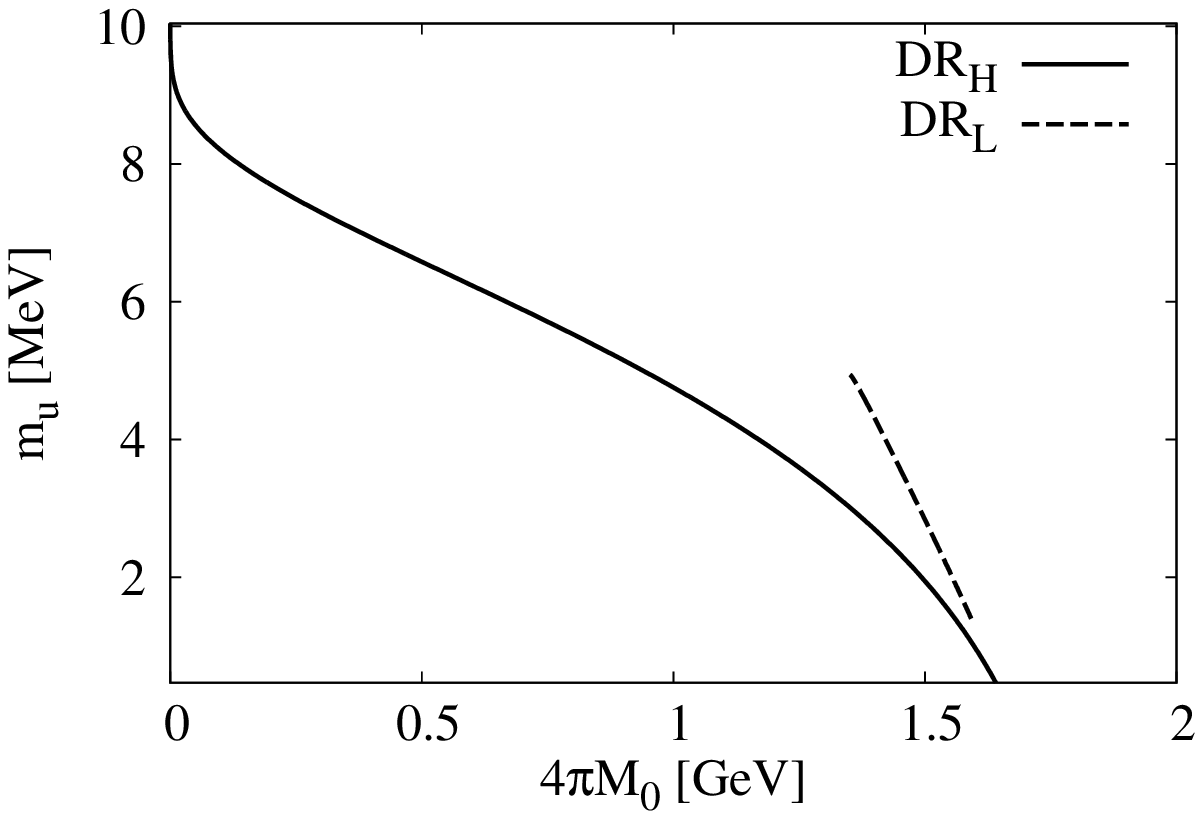}
   \includegraphics[width=7.5cm,keepaspectratio]{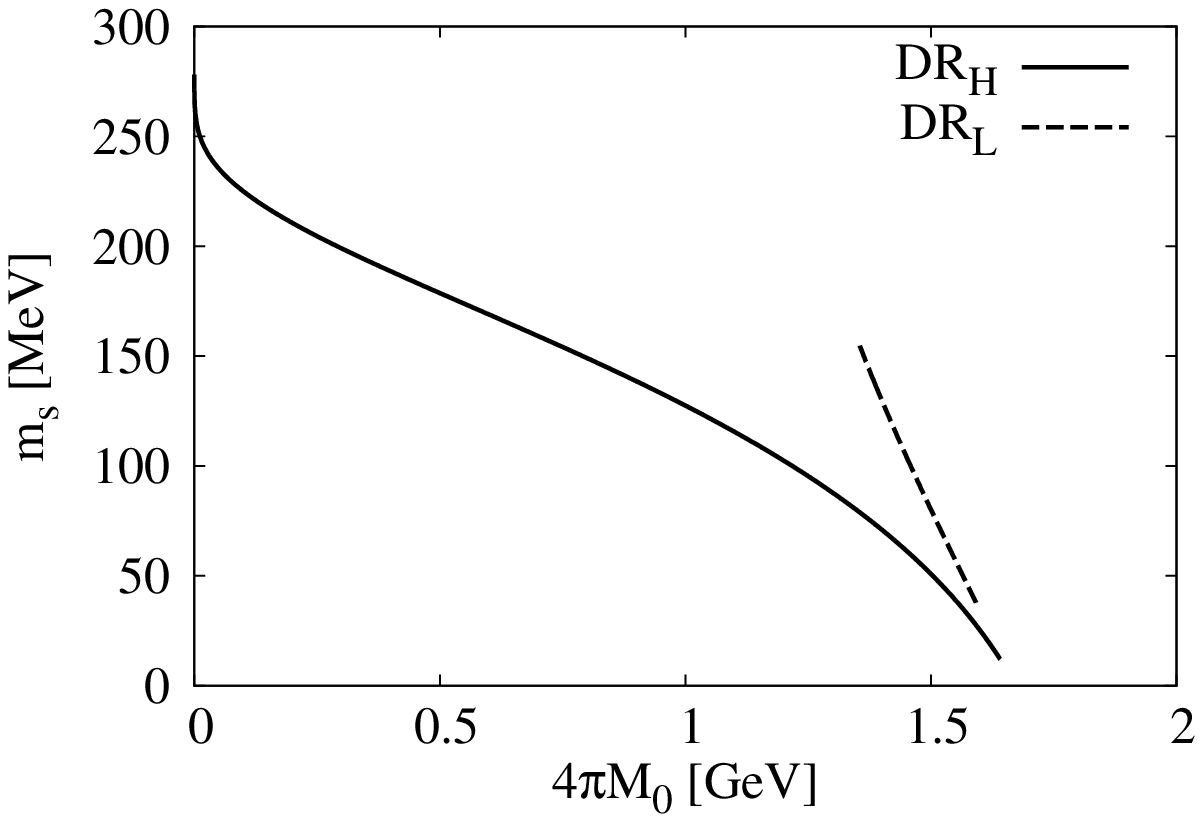}
   \includegraphics[width=7.5cm,keepaspectratio]{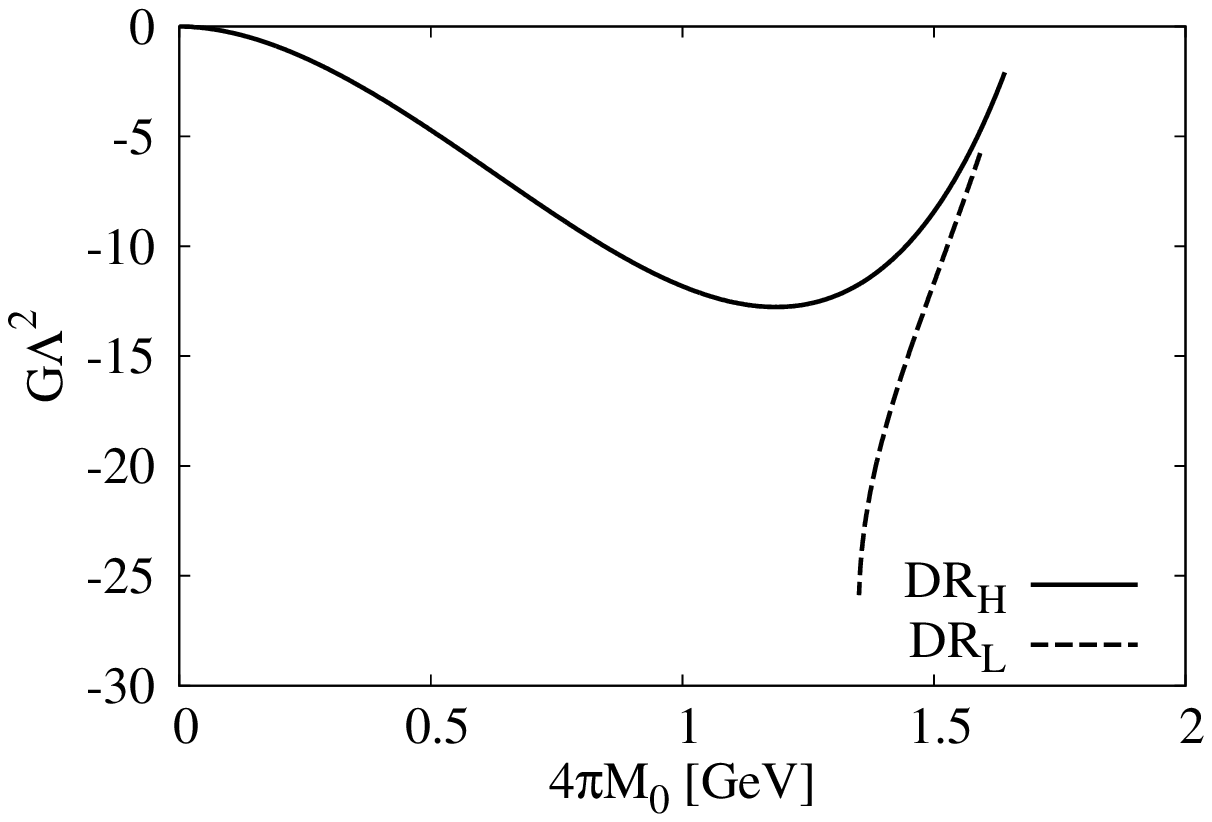}
   \includegraphics[width=7.5cm,keepaspectratio]{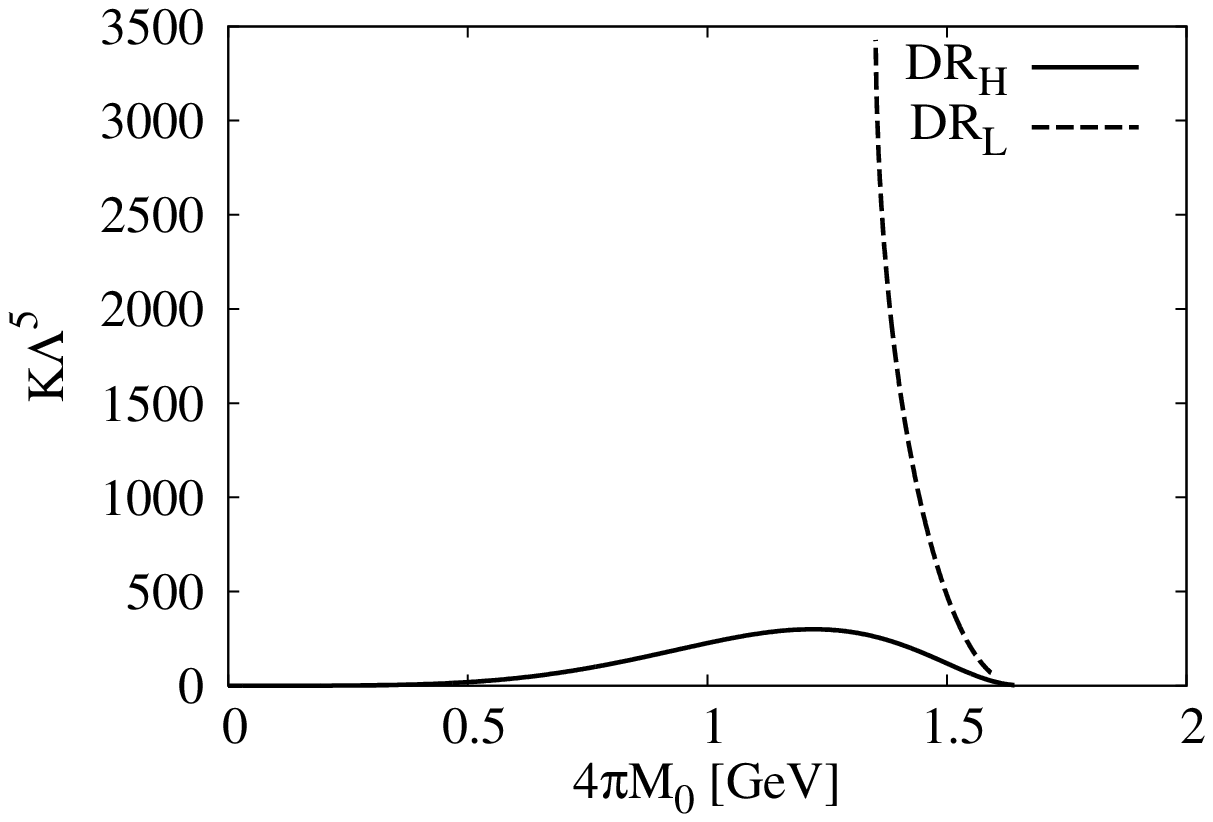}
   \includegraphics[width=7.5cm,keepaspectratio]{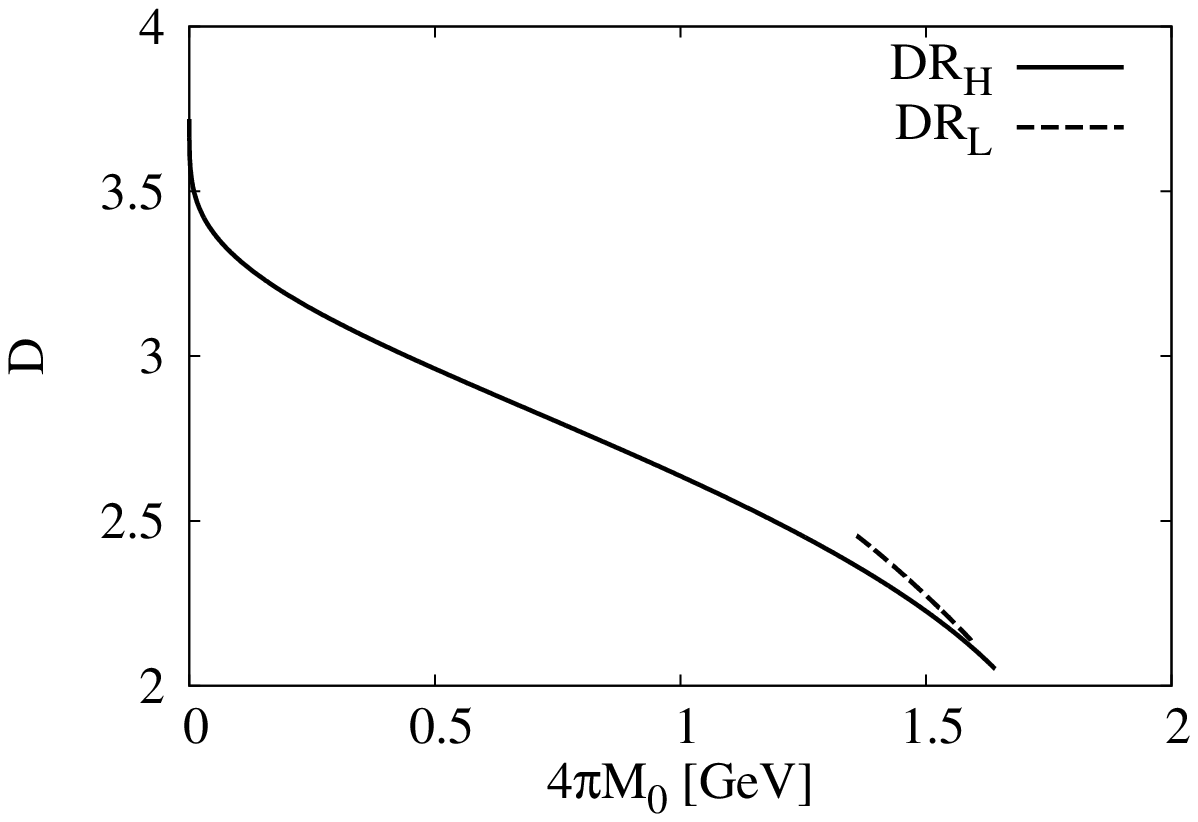}
   \caption{\label{fig:para_DR}
     Parameters in the DR.}
\end{center}
\end{figure}
We align the resulting parameters for the DR case in Fig.~\ref{fig:para_DR}
where we chose the horizontal axis to be $4 \pi M_0$ as mentioned above.
It should be noted that the DR has two sets of solutions under the
conditions discussed above. As is seen in the figure, we indicate the
solutions with higher and the lower dimensions by denoting
${\rm DR_H}$ and ${\rm DR_L}$, respectively.
One sees that the current quark masses, $m_u$ and $m_s$, decrease
according to $\Lambda_{\rm DR}$ in the DR method, whose tendencies
and the values are similar to the cases with the other regularizations.
Note that the increase of $\Lambda_{\rm DR}(=4\pi M_0)$ means
the decrease of $D$ as seen in Fig.~\ref{fig:para_mu}. While the
couplings $G$ and $K$ show non-monotonic curves for the higher
dimensional case, ${\rm DR_H}$, which is distinguishing feature of the
${\rm DR_H}$. We will consider the cause on this
non-monotonic behavior of the couplings in Sec.~\ref{sec:discussion}.
It is noted that for each parameter, the existing region of the ${\rm DR_L}$
case is considerably narrow compare to the ${\rm DR_H}$.

It is also interesting to mention that the parameters exist for considerably
higher scale up to several GeV; where we do not show upper limit
since the models are no longer reliable for such high scale.
While the lower limits are important because they are comparable to
hadronic scale and it has a strong regularization dependence.
Here we numerically find the following values for lower limits:
\begin{eqnarray}
  {\rm 3D:}\,\,\, 580.5{\rm MeV}, \quad
  {\rm 4D:}\,\,\, 719.3{\rm MeV}, \quad
  {\rm PV:}\,\,\, 717.7{\rm MeV}, \quad
  {\rm PT:}\,\,\, 629.0{\rm MeV}.
\end{eqnarray}

On the other hand, there does not appear the lower limit in the DR case;
the curves can be drawn in the $M_0 \to 0$ limit. It may also be interesting
to note that we have the upper limit in the DR. As is seen in the lower panel
of Fig.~\ref{fig:para_DR}, the value is around $1.7$GeV. This limit stems
from the restriction on the dimension, $2 < D <4$.

\subsection{\label{subsec:Para_Table}%
Table of Parameters}
We have drawn the tendency of the parameters according to $\Lambda$
in Sec. \ref{subsec:para_L}. We now align the values of parameters in the
table form. It will be useful for the study of the meson properties and the
various physical phenomena such as chiral phase transition based on the
NJL model. We also align the parameter sets obtained in the preceding
analyses for various regularization methods, then make the numerical
comparison among them.

\begin{table}[h!]
\caption{\label{table_3D_para}
  Parameters and $m^*$ in the 3D cutoff. }
\begin{center}
\begin{tabular}{ccccc|cc|c}
\hline
$\Lambda_{\rm 3D}$(MeV) & $m_u$(MeV)  & $m_s$(MeV)
 & $G \Lambda^2$  & $K \Lambda^5$ & $m_u^*$(MeV)& $m_s^*$(MeV)& Ref.  \\
\hline
960.4 & 3.00 & 89.45 & 1.552 & 8.339 & 212 & 417 & \\
681.6 & 5.00 & 128.3  & 1.706 & 8.772 & 286 & 487 & \\
630.9 & 5.50 & 135.9 & 1.814 & 9.165 & 324 & 519 & \\
625.4 & 5.55 & 136.6 & 1.830 & 9.246 & 330 & 524 & \\
580.5 & 5.87 & 139.1 & 2.087 & 10.08 & 414 & 592 & \\
\hline
$631.4$ & $5.5$ & $135.7$ & $1.835$ & $9.29$& 335 & 527 &  \cite{Hatsuda:1994pi} \\
$602.3$ & $5.5$ & $140.7$ & $1.835$ & $12.36$ & 367.7 & 549.5 & \cite{Rehberg:1995kh} \\
\hline
\end{tabular}
\end{center}
\end{table}
\begin{table}[h!]
\caption{\label{table_4D_para}
  Parameters and $m^*$ in the 4D cutoff. }
\begin{center}
\begin{tabular}{ccccc|cc|c}
\hline
$\Lambda_{\rm 4D}$(MeV) & $m_u$(MeV)  & $m_s$(MeV)
 & $G \Lambda^2$  & $K \Lambda^5$ & $m_u^*$(MeV)& $m_s^*$(MeV)& Ref. \\
\hline
1421 & 3.00 & 94.41 & 3.026 & 55.02 & 191 & 400 & \\
1046 & 5.00 & 139.1 & 3.156 & 57.80 & 228 & 445 &  \\
850.2 & 7.00 & 176.5 & 3.526 & 64.30 & 280 & 499 &  \\
772.4 & 8.14 & 194.8 & 3.990 & 72.60 & 330 & 545 &  \\
719.3 & 8.99 & 205.5 & 5.341 & 91.99 & 453  & 648 & \\
\hline
1049 & 5.0 & --  & 3.741 & -- & 222.3 & --  &  \cite{Krewald:1991tz} (2f.) \\
854 & 7.0 & --  & 4.230 & -- & 270.9 & --  &  \cite{Krewald:1991tz} (2f.) \\
\hline
\end{tabular}
\end{center}
\end{table}
\begin{table}[h!]
\caption{\label{table_PV_para}
  Parameters and $m^*$ in the PV. }
\begin{center}
\begin{tabular}{ccccc|cc|c}
\hline
$\Lambda_{\rm PV}$(MeV) & $m_u$(MeV)  & $m_s$(MeV)
 & $G \Lambda^2$  & $K \Lambda^5$ & $m_u^*$(MeV)& $m_s^*$(MeV)& Ref.  \\
\hline
1443 & 3.00 & 102.7 & 3.094 & 59.83 & 188 & 385 & \\
1085 & 5.00 & 158.0 & 3.305 & 68.29 & 218 & 420 & \\
910.9 & 7.00 & 210.5 & 3.705 & 85.61 & 248 & 454 & \\
743.3 & 11.8 & 327.8 & 5.885 & 175.5 & 330 & 534 & \\
717.7 & 15.6 & 396.6 & 9.282 & 310.3 & 404 & 585 & \\
\hline
$1400$ & $2.7$ & $92$ & $ 3.038$ & $473.3$ & 214 & 397&  \cite{Osipov:2004bj} \\
$980$ & $4.7$ & $155$ & $ 3.457$ & $431.2$ & 286 & 485&  \cite{Osipov:2004bj} \\
\hline
\end{tabular}
\end{center}
\end{table}
\begin{table}[h!]
\caption{\label{table_PT_para}
  Parameters and $m^*$ in the PT. }
\begin{center}
\begin{tabular}{ccccc|cc|c}
\hline
$\Lambda_{\rm PT}$(MeV) & $m_u$(MeV)  & $m_s$(MeV)
 & $G \Lambda^2$  & $K \Lambda^5$ &  $m_u^*$(MeV)& $m_s^*$(MeV)& Ref. \\
\hline
1489 & 3.00 & 100.2 & 2.926 & 68.03 & 172 & 383 & \\
1115 & 5.00 & 150.4 & 2.941 & 74.97 & 194 & 418 & \\
924.1 & 7.00 & 195.6 & 3.059 & 85.50 & 216 & 451 & \\
650.9 & 14.5 & 338.6 & 5.169 & 159.9 & 330 & 593 & \\
629.0 & 17.2 & 380.9 & 7.493 & 222.4 & 415 & 665 & \\
\hline
1080 & 5.0 & --  & 3.802 & -- & 216  & -- &  \cite{Cui:2014hya}(2f.) \\
907 & 7.0  & -- & 4.138 &  -- & 240  & -- &  \cite{Cui:2014hya}(2f.) \\
\hline
\end{tabular}
\end{center}
\end{table}
\begin{table}[h!]
\caption{\label{table_DRH_para}
  Parameters and $m^*$ in the ${\rm DR_H}$. }
\begin{center}
\begin{tabular}{cccccc|cc|c}
\hline
$\Lambda_{\rm DR}$(MeV) & $D$ & $m_u$(MeV)  & $m_s$(MeV)
 & $G \Lambda^{D-2}$  & $K \Lambda^{2D-3}$  &  $m_u^*$(MeV)& $m_s^*$(MeV)& Ref.  \\
\hline
1353 & 2.368 & 3.00 & 79.23 & $-0.1880$ & 0.06658 & $-591$ & $-680$ & \\
1170 & 2.515 & 4.00 & 106.6 & $-0.2977$ & 0.1604 & $-621$ & $-706$ & \\
935.5 & 2.677 & 5.00 & 134.3 & $-0.3930$ & 0.2371 & $-653$ & $-734$ & \\
667.8 & 2.853 & 6.00 & 162.3 & $-0.4203$ & 0.1814 & $-685$ & $-763$ & \\
380.1 & 3.044 & 7.00 & 190.6 & $-0.2659$ & 0.04530 & $-719$ & $-793$ & \\
\hline
$1382$ & $2.37$ & $3.0$ & -- & $-0.1647$ & -- & $-570$ & -- & \cite{Inagaki:2015lma} (2f.) \\
$1219$ & $2.56$ & $5.0$ & -- & $-0.3144$ & -- & $-519$ & -- & \cite{Inagaki:2015lma} (2f.) \\
\hline
\end{tabular}
\end{center}
\end{table}
\begin{table}[h!]
\caption{\label{table_DRL_para}
  Parameters and $m^*$ in the ${\rm DR_L}$. }
\begin{center}
\begin{tabular}{cccccc|cc|c}
\hline
$\Lambda_{\rm DR}$(MeV) & $D$ & $m_u$(MeV)  & $m_s$(MeV)
 & $G \Lambda^{D-2}$  & $K \Lambda^{2D-3}$  &  $m_u^*$(MeV)& $m_s^*$(MeV)& Ref.  \\
\hline
1489 & 2.289 & 3.00 & 85.26 & $-0.1627$ & 0.09688 & $-467$ & $-546$ & \\
1421 & 2.379 & 4.00 & 118.6 & $-0.2792$ & 0.3459 & $-461$ & $-529$ & \\
1351 & 2.463 & 4.94 & 157.1 & $-0.5289$ & 1.433 & $-456$ & $-500$ & \\
\hline
$1382$ & $2.37$ & $3.0$ & -- & $-0.1647$ & -- & $-570$ & -- & \cite{Inagaki:2015lma} (2f.) \\
$1219$ & $2.56$ & $5.0$ & -- & $-0.3144$ & -- & $-519$ & -- & \cite{Inagaki:2015lma} (2f.) \\
\hline
\end{tabular}
\end{center}
\end{table}
In Tab. \ref{table_3D_para},
our result is almost coincident with the one in Ref.~\cite{Hatsuda:1994pi}.
The slight difference comes from the input parameter. The difference of
$K \Lambda^5$ between our result and Ref.  \cite{Rehberg:1995kh} is caused
by the definition of the $\eta'$ mass, the former only considers the real
part of the
propagator, while the latter includes the imaginary part.
In the 4D and PT cases, our results are close to the previous ones which obtained
by two-flavor NJL model. Ref.~\cite{Krewald:1991tz}
uses the decay width of $\pi^0 \to \gamma \gamma$ as the input parameter.
In Tab.~\ref{table_PV_para}, our result is close to the previous one
\cite{Osipov:2004bj} except for the value of $K \Lambda^5$. To obtain the
meson mass spectra, the generalized heat kernel expansion  are used in
Ref.~\cite{Osipov:2004bj}. Although there exists one set of solution
in the two-flavor model, two sets of solutions appear in the three-flavor case.
As decreasing the cutoff scale $\Lambda_{\rm DR}$, $m_u^*$ and $m_s^*$
increase in the ${\rm DR_H}$ and decrease in the ${\rm DR_L}$.

\section{\label{sec:prediction}%
Model predictions}
The model parameters have been carefully fitted in the previous section.
We are now ready for investigating various predicted quantities;
in this section we are going to analyze the chiral condensates and constituent
quark masses, the meson properties and the topological susceptibility. As is
well known, the model predictions depend on the regularization methods
and the model parameters, whose test is the focus we will study here.

\subsection{\label{subsec:PM}%
Chiral condensates and constituent quark masses}
The chiral condensates are the key quantities in this kind of model, since
they are intimately related to the chiral symmetry breaking of the system and
critically determine the model behavior. The constituent quark masses are
also important physical objects which explicitly appear in the equations
for the meson properties as shown in the Appendix. Since the mass of the
proton and neutron are around $1$GeV, then we naively expect that the
values of the constituent masses for up and down quarks are around
$1/3$GeV, i.e., $330$MeV. We will confirm that this value is consistently
achieved within reliable model scale in the current models.

\begin{figure}[h!]
\begin{center}
   \includegraphics[width=7.5cm,keepaspectratio]{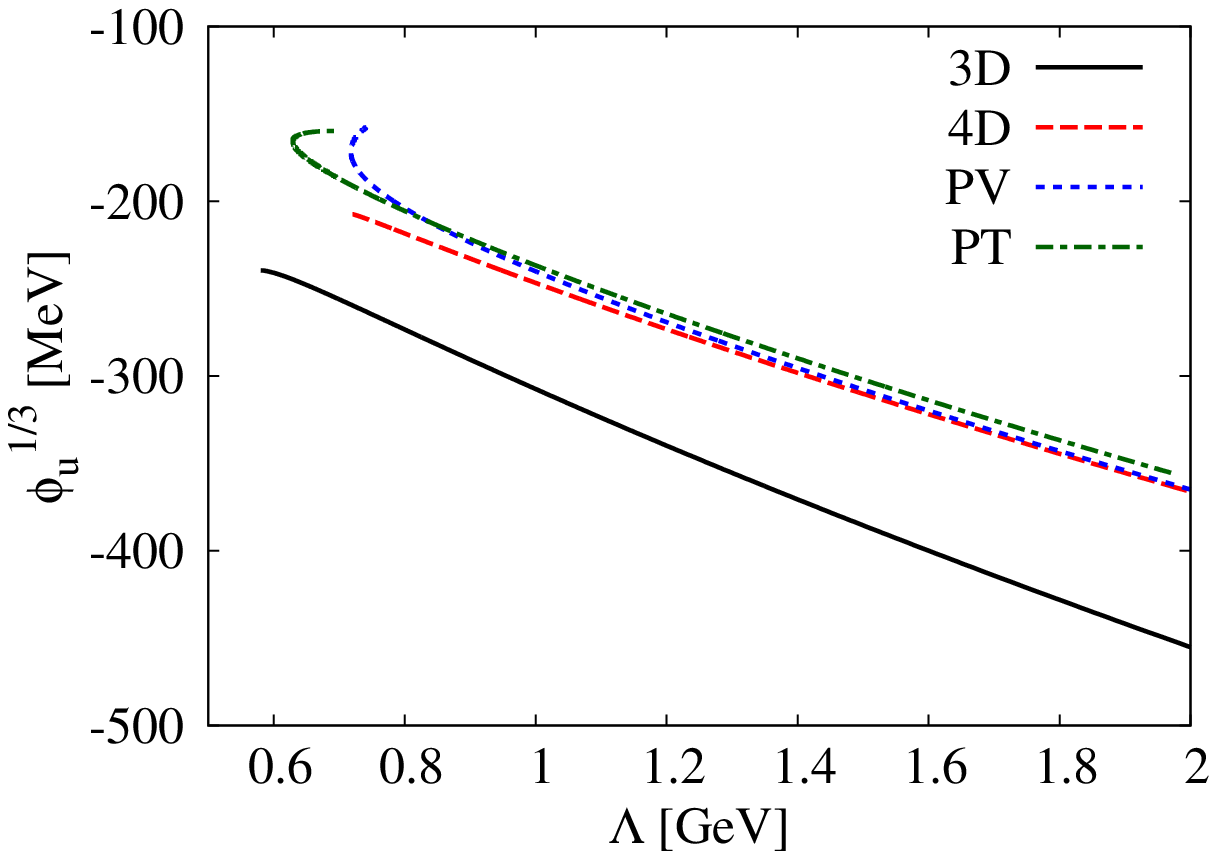}
   \includegraphics[width=7.5cm,keepaspectratio]{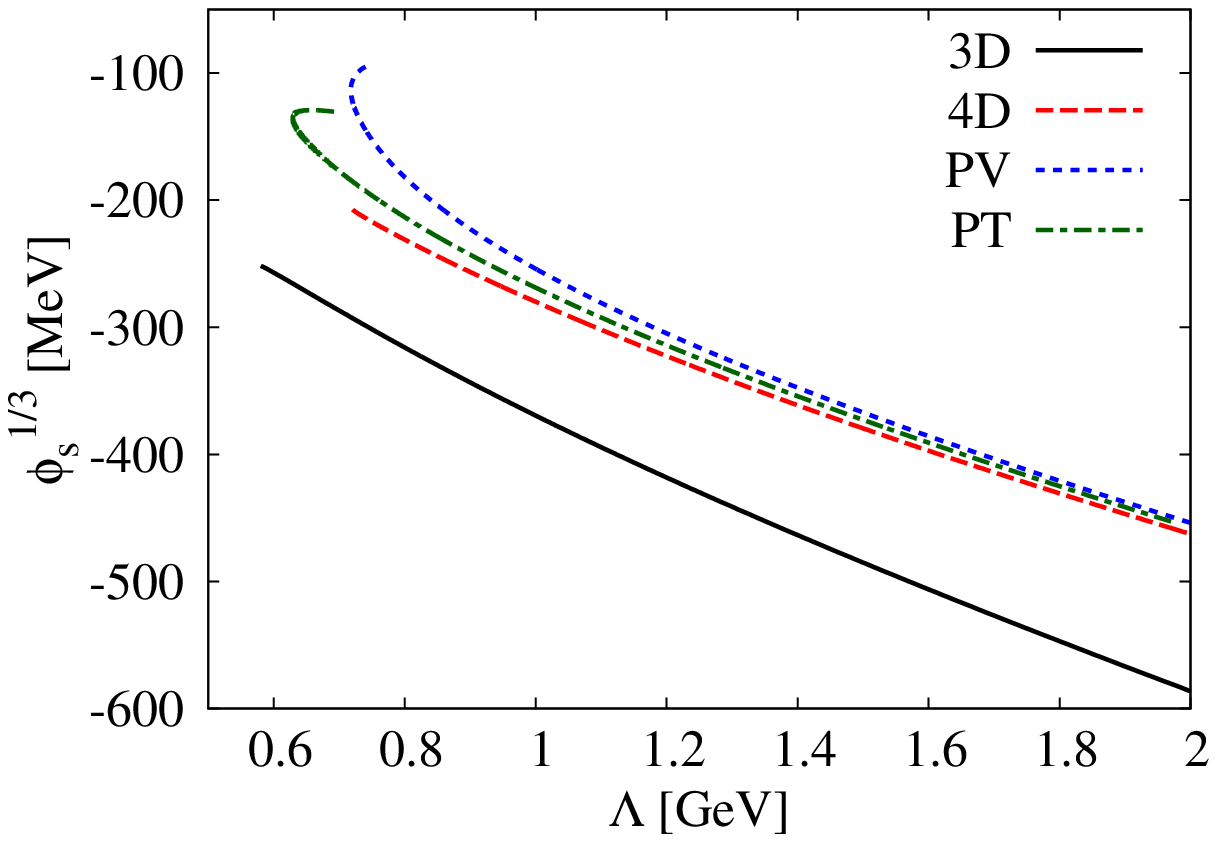}
   \includegraphics[width=7.5cm,keepaspectratio]{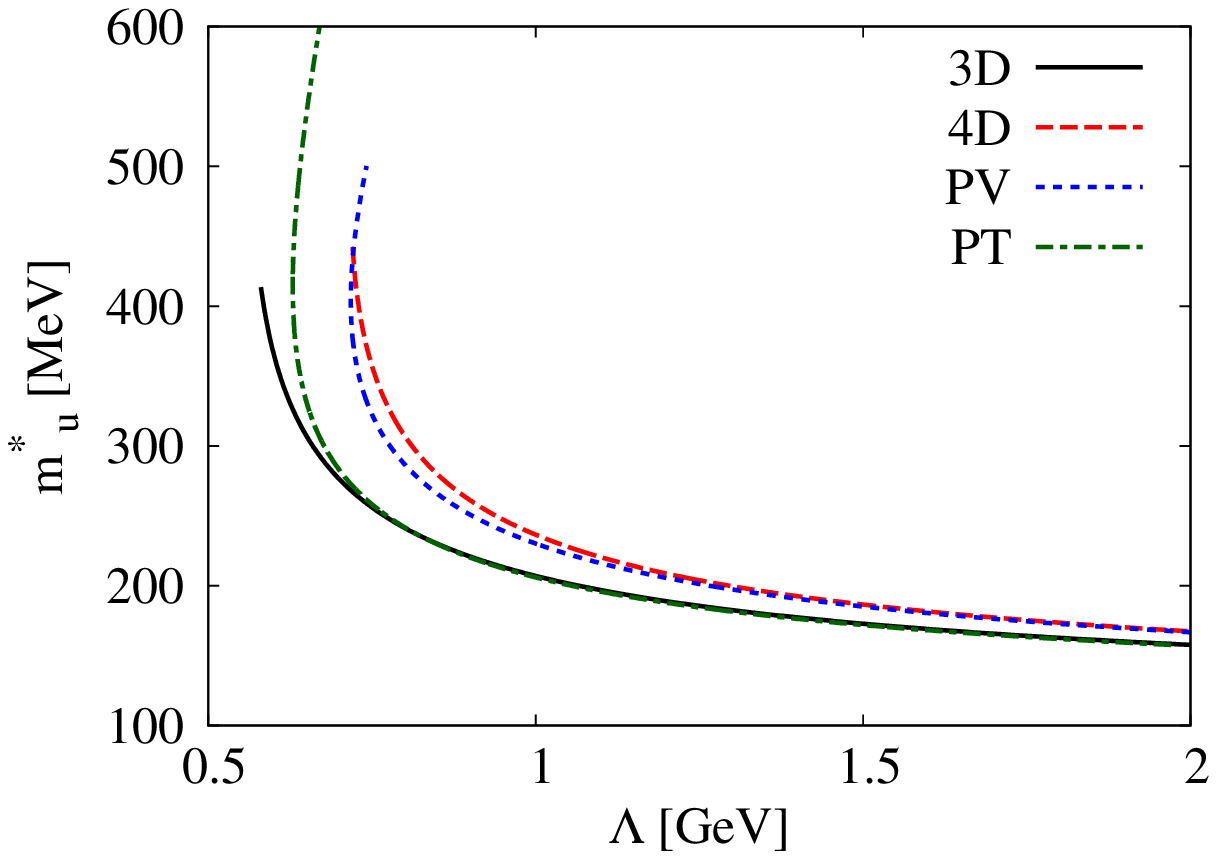}
   \includegraphics[width=7.5cm,keepaspectratio]{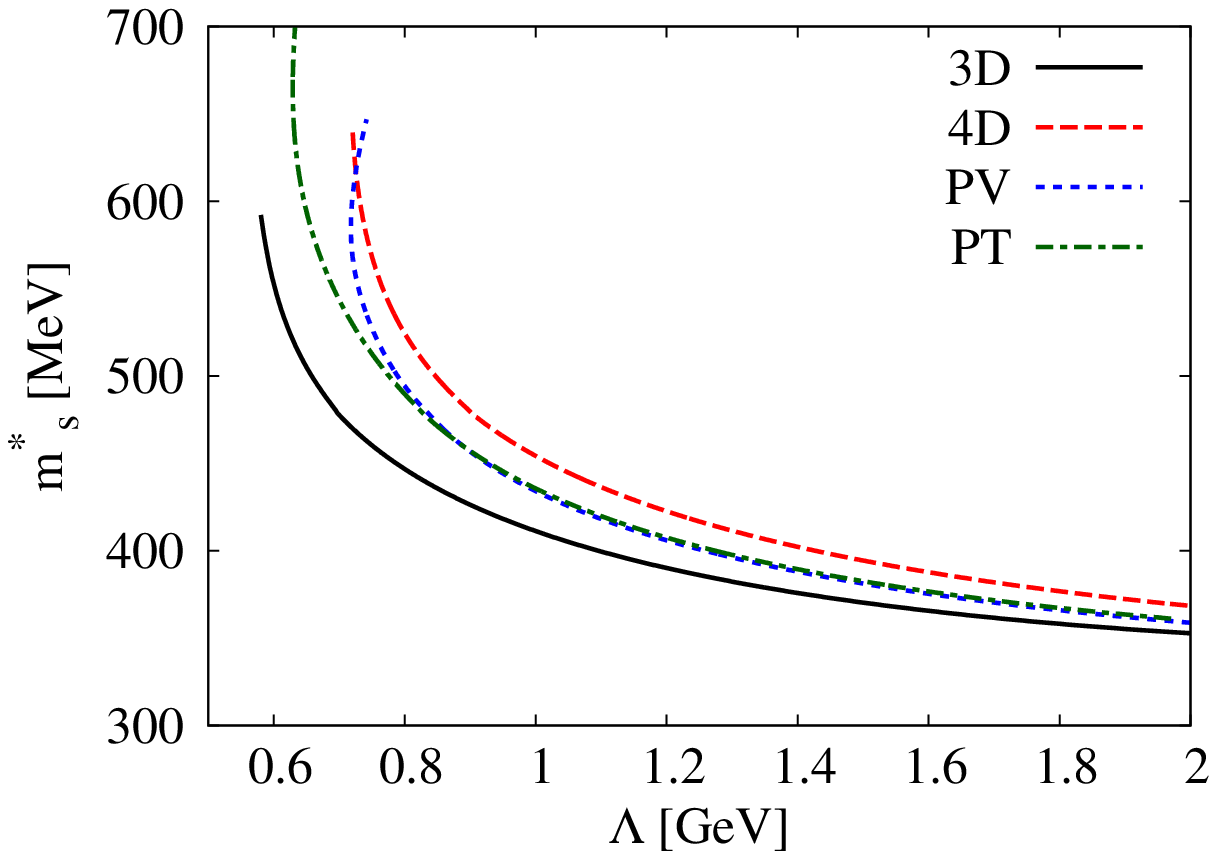}
   \caption{\label{fig:PM}
      $\phi_i$ and $m^*_i$ in the 3D, 4D, PV, PT.}
\end{center}
\end{figure}
Figure \ref{fig:PM} displays the results of the chiral condensates $\phi_i$
and the constituent quark masses $m^*_i$ with the fitted parameters shown
in the previous subsection. One notes that the absolute values of $\phi_i$
monotonically increase with the cutoff, while $m_i^*$ decrease with
respect to $\Lambda$. Observing the obtained values, we see that the
regions for $0.6 \lesssim \Lambda \lesssim 1$GeV have nice numerical
plots, $\phi_u^{1/3} \simeq -230$MeV and $m_u^* \simeq 330$MeV.
One also notes that
the  constituent quark masses $m_i^*$ become nearly constant for high
cutoff scale in all regularization cases. This characteristic may show us
some universal properties that the constituent quark masses possess,
which we will discuss in more detail in Sec.~\ref{subsec:high_L}.

\begin{figure}[h!]
\begin{center}
   \includegraphics[width=7.5cm,keepaspectratio]{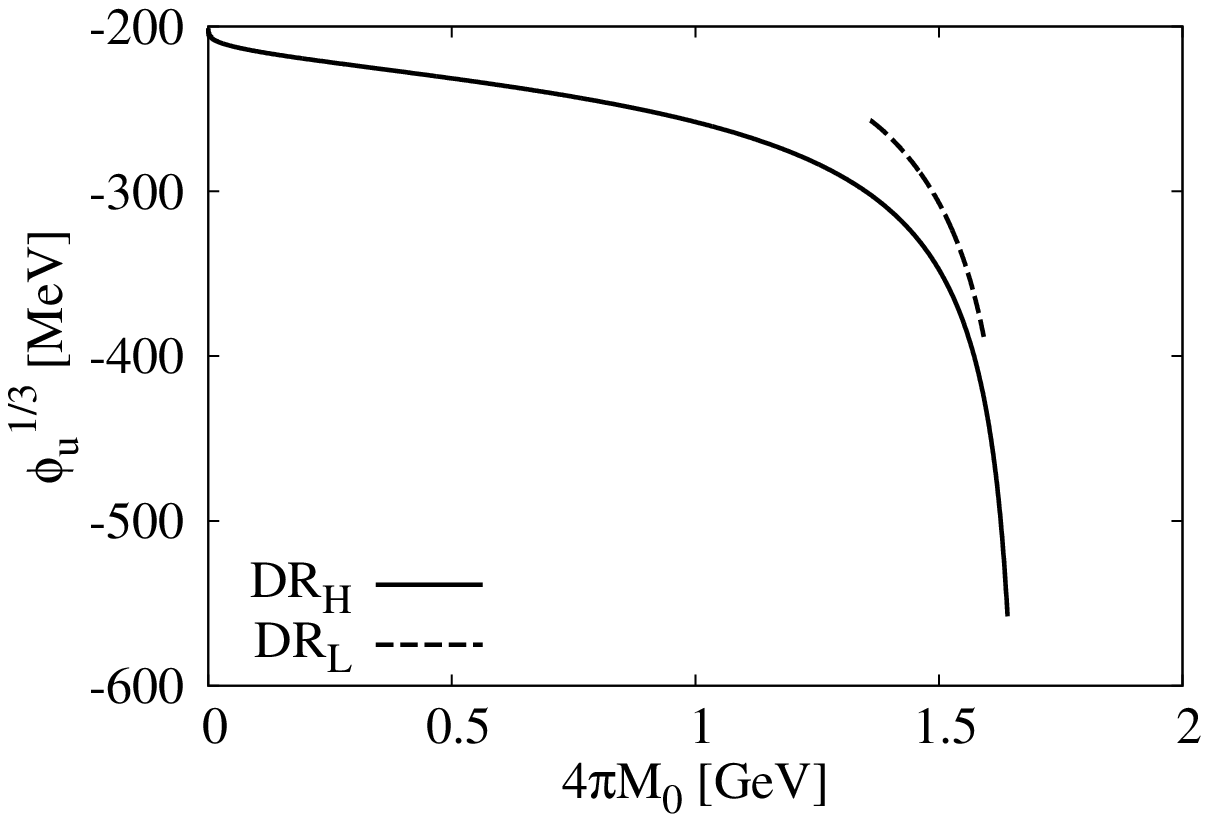}
   \includegraphics[width=7.5cm,keepaspectratio]{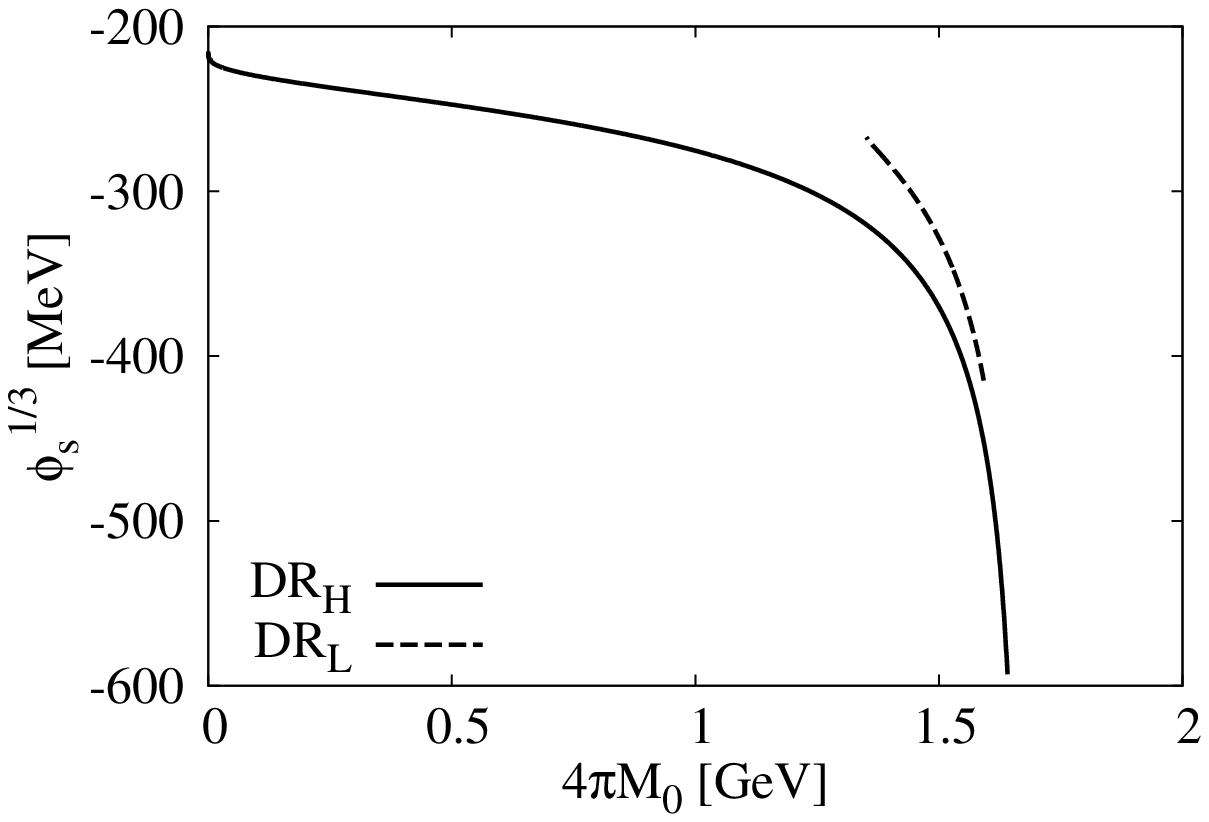}
   \includegraphics[width=7.5cm,keepaspectratio]{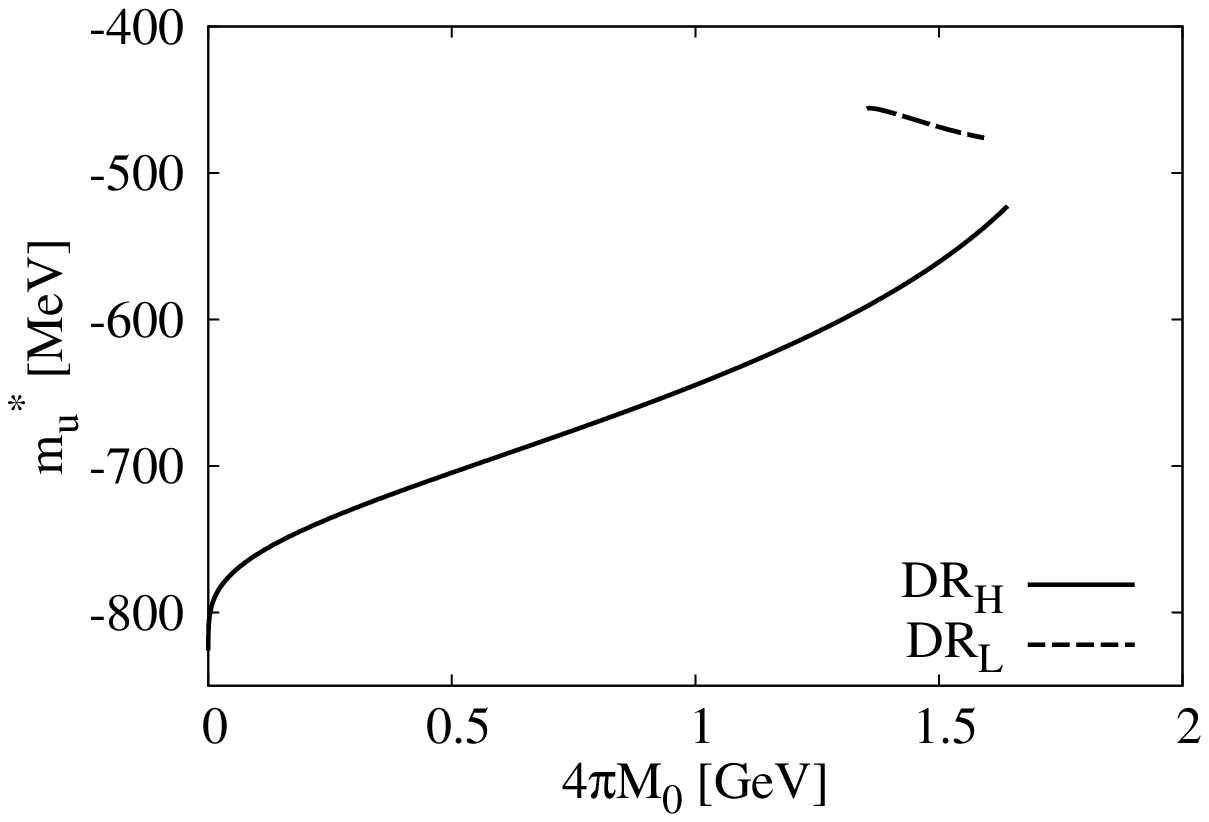}
   \includegraphics[width=7.5cm,keepaspectratio]{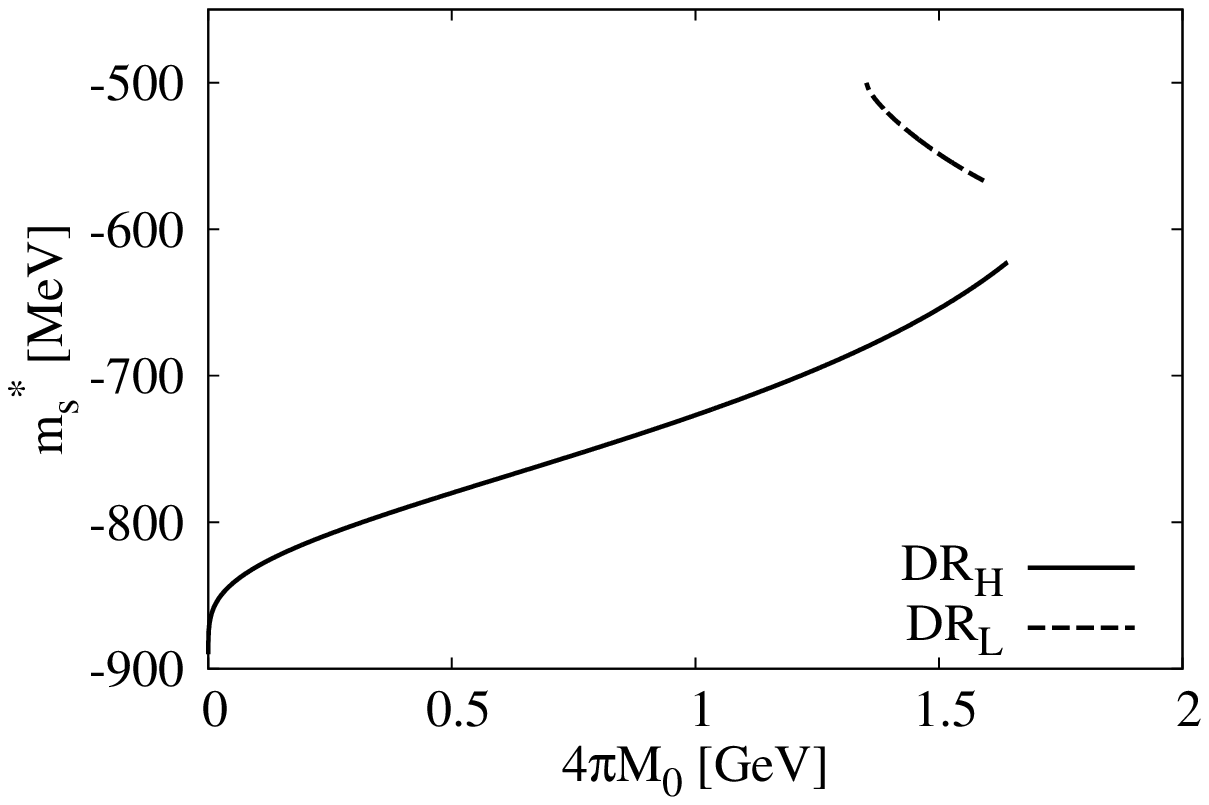}
   \caption{\label{fig:PM_DR}
        $\phi_i$ and $m^*_i$ in the DR. 
      }
\end{center}
\end{figure}
We exhibit the resulting chiral condensates and constituent masses in the
DR method in Fig.~\ref{fig:PM_DR}. The qualitative tendency of the chiral
condensates are similar to the other regularization ways as compared
with the upper two panels of Fig.~\ref{fig:PM} where the absolute values
of $\phi_i$ become larger with increasing the model scale. On the other
hand, for the constituent quark masses, one notices the crucial difference
between the DR and the other prescriptions; the signs of the constituent
quark masses are opposite. This comes from the mathematical treatment
of the analytic continuation when we regularize the loop integrals.
The sign of the constituent quark masses can be positive with the
counter terms~\cite{Inagaki:2012re}. The absolute values of the
constituent quark masses decrease according to the model scale, which
is the case for all the regularizations.

\subsection{\label{subsec:pre_meson}%
Meson properties and topological susceptibility}
In this subsection we calculate the predicted meson properties,
the $\eta$ meson mass~$m_\eta$, the kaon decay constant~$f_{\rm K}$,
the sigma meson mass~$m_\sigma$ and the topological susceptibility~$\chi$
through using the fixed parameters. It is practically interesting, since the
numerics are effectively important in determining whether the model
and the employed methods including the choice of parameters are appropriate,
eventually to be tested by experiments.

\begin{figure}[h!]
\begin{center}
   \includegraphics[width=7.5cm,keepaspectratio]{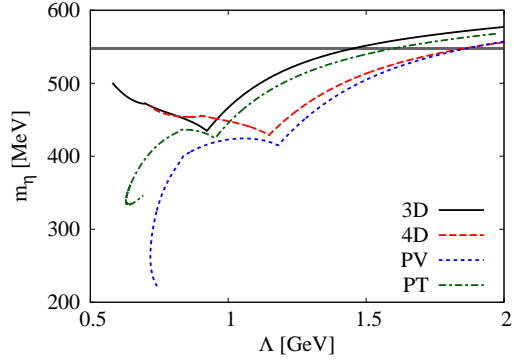}
   \caption{\label{fig:eta}
     $m_\eta$ in the 3D, 4D, PV, PT. The gray line indicates the experimental value,
     $548$MeV.}
\end{center}
\end{figure}
Figure~\ref{fig:eta} displays the obtained results for the $\eta$ mass $m_\eta$
in each regularization. Since the $\eta$ mass is one of the input quantities used
for the fitting in the DR, it is fixed at $548$GeV. We see that the obtained values
are smaller than the experimental data for $\Lambda$ being the hadronic scale,
while the experimental line crosses for relatively larger $\Lambda$.
The curves for the 3D  and 4D indicate similar structures; they decrease upto
some $\Lambda$, then turn to increase for large scale. The curves in the PV and
PT cases show almost monotonic behavior; they become larger with increasing
$\Lambda$. In any regularization, there appears typical cusp around
$\Lambda \sim 1$GeV, which comes from the complex property due to the
determinantal form of the equation (\ref{eq:eta}) for $\eta^{\prime}$ and $\eta$
masses.

\begin{figure}[h!]
\begin{center}
   \includegraphics[width=7.5cm,keepaspectratio]{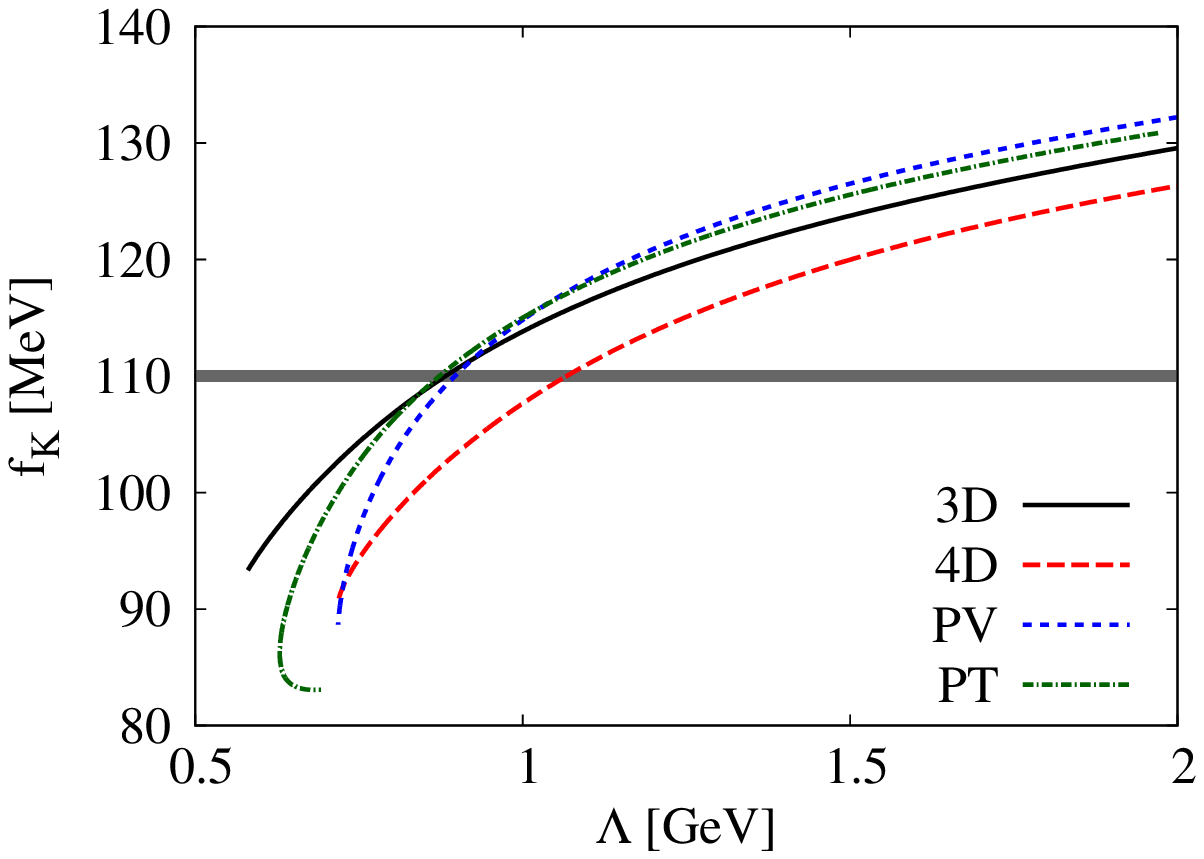}
   \includegraphics[width=7.5cm,keepaspectratio]{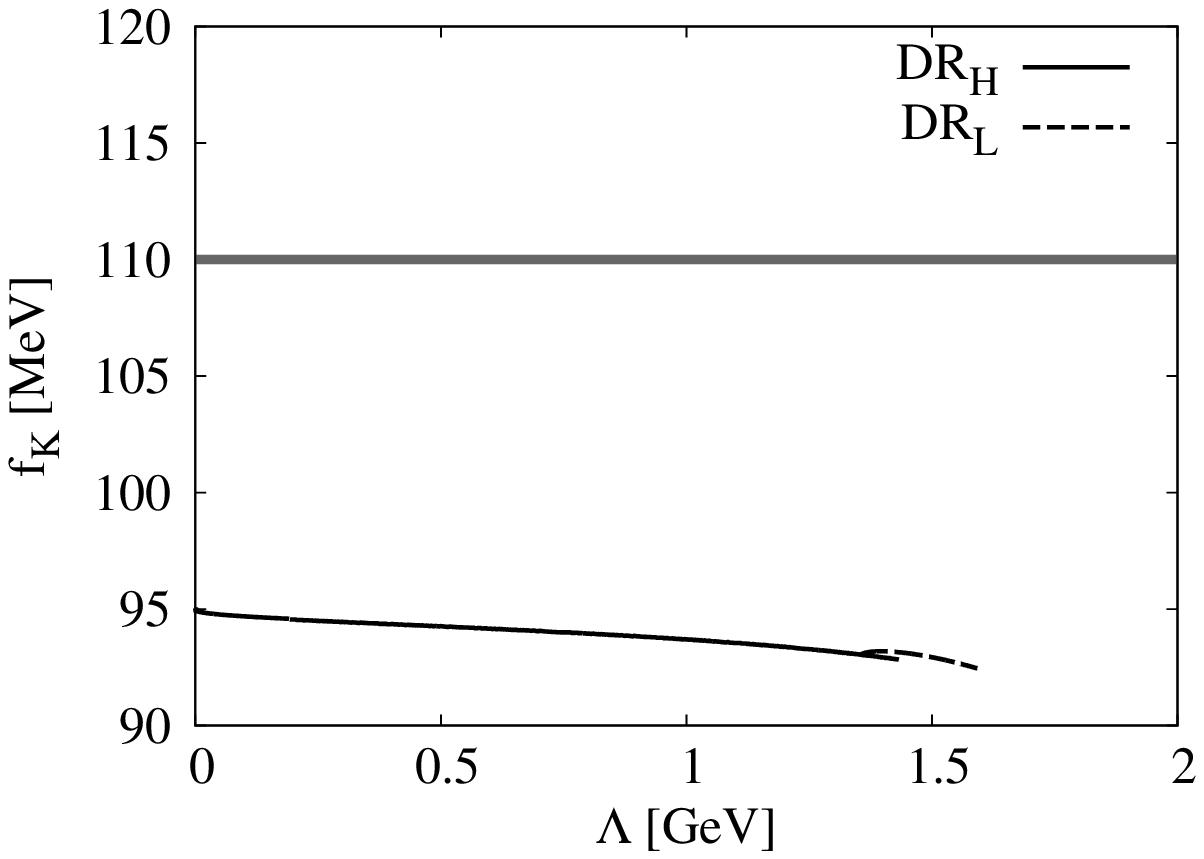}
   \caption{\label{fig:f_K}
     $f_{\rm K}$ in the 3D, 4D, PV, PT (left) and the DR (right).}
\end{center}
\end{figure}
The kaon decay constants are shown in Fig.~\ref{fig:f_K},  where the results
for the 3D, 4D, PV and PT all increase with the cutoff scale (left panel), while
it becomes smaller according to $4\pi M_0$ in the DR. 
In $0.8 \lesssim \Lambda \lesssim 2$GeV the curves for the 3D, PV and PT  
have similar behavior, while the curve for 4D shows a smaller $f_K$ than the others.
It is interesting that the results for 3D, PT and PT have close value to the
experimental data at almost same cutoff scale, $\Lambda \simeq 0.9$GeV.

\begin{figure}[h!]
\begin{center}
   \includegraphics[width=7.5cm,keepaspectratio]{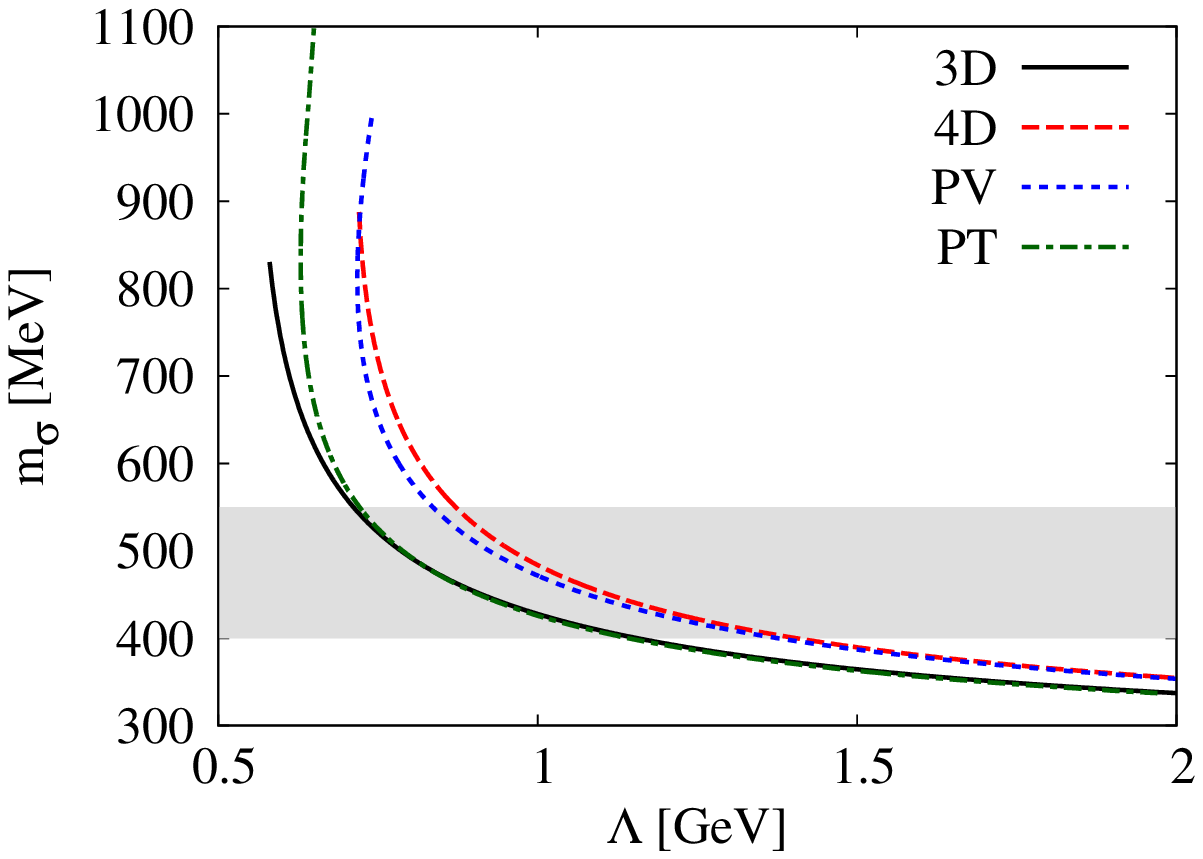}
   \includegraphics[width=7.5cm,keepaspectratio]{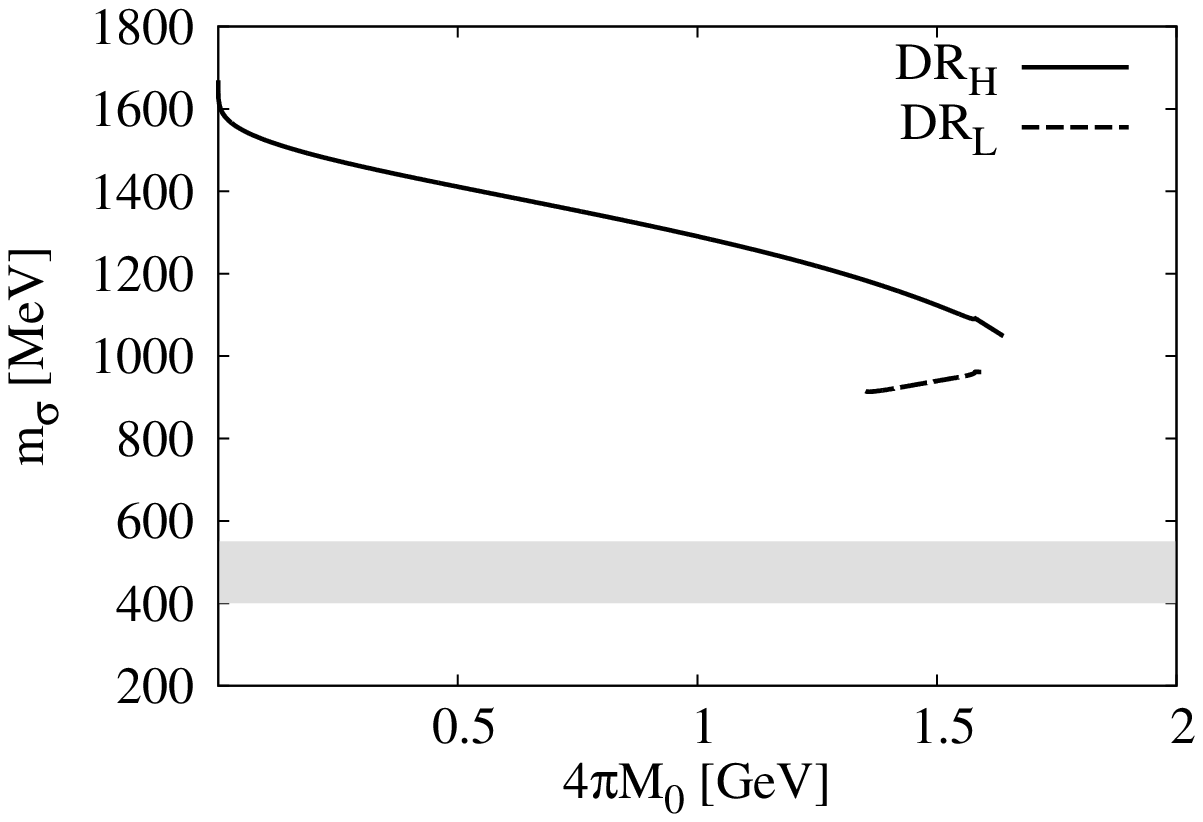}
   \caption{\label{fig:sigma}
        $m_\sigma$ in the 3D, 4D, PV, PT (left) and the DR (right).
        The gray regions show the empirical range~\cite{Agashe:2014kda}.}
\end{center}
\end{figure}
The sigma meson mass, $m_\sigma$, is evaluated as a pole of the propagator
calculated from the loop integral of the scalar channel calculated by
Eq.~(\ref{eq:scalar}), whose results are exhibited in Fig.~\ref{fig:sigma}.
The numbers obtained in the four regularizations
observed in the left panel have reliable range compare to the empirical
value, $m_\sigma \simeq 400-550$MeV~\cite{Agashe:2014kda},
while the DR curves read the considerably larger values than the
empirical scale. In the DR the pole of the sigma meson propagator
does not exist for $\Lambda \gtrsim 1.5$GeV, we regard the maximum value as
the sigma meson mass in such region \cite{inagaki:2008}.
Contrary to the case for
the $\eta$ mass, the mass decreases monotonically with respect to the
model scale, which stems from the monotonic function of the equation
for the sigma meson.

\begin{figure}[h!]
\begin{center}
   \includegraphics[width=7.5cm,keepaspectratio]{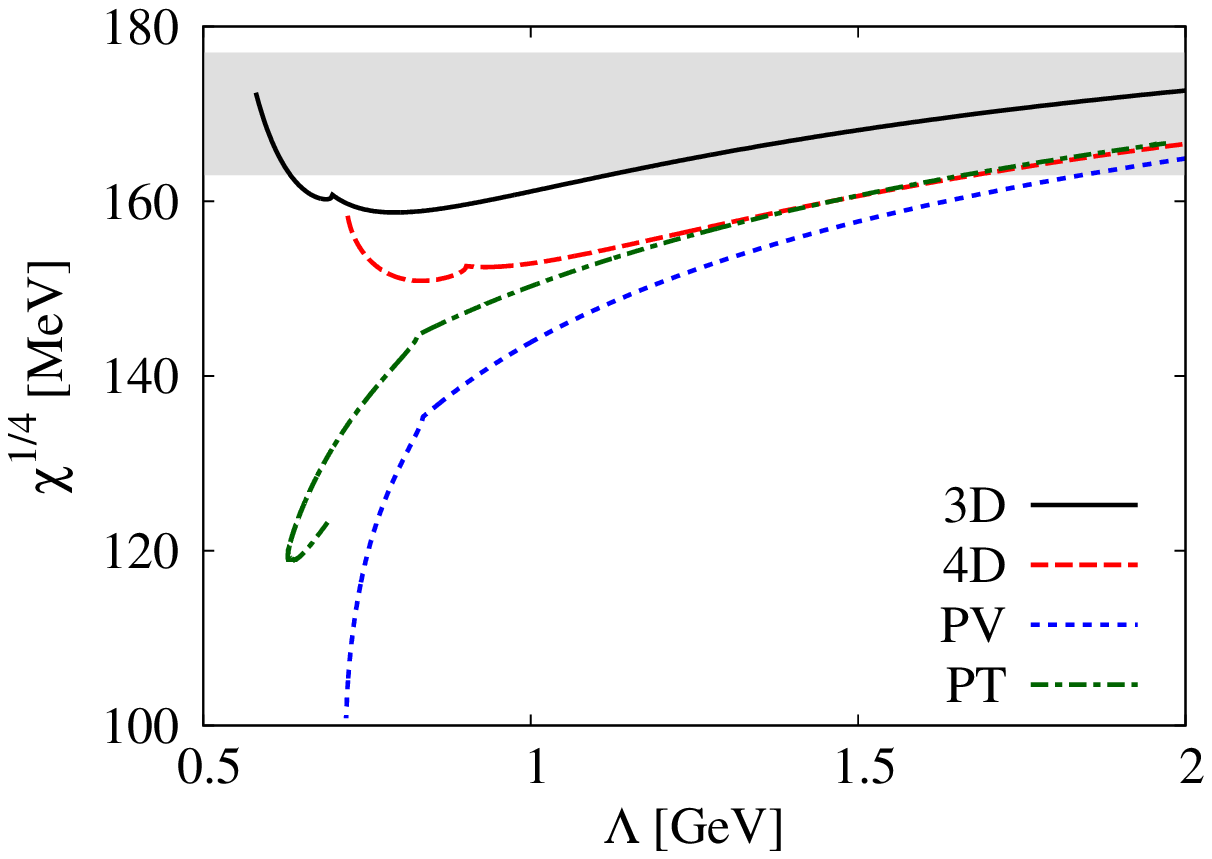}
   \includegraphics[width=7.5cm,keepaspectratio]{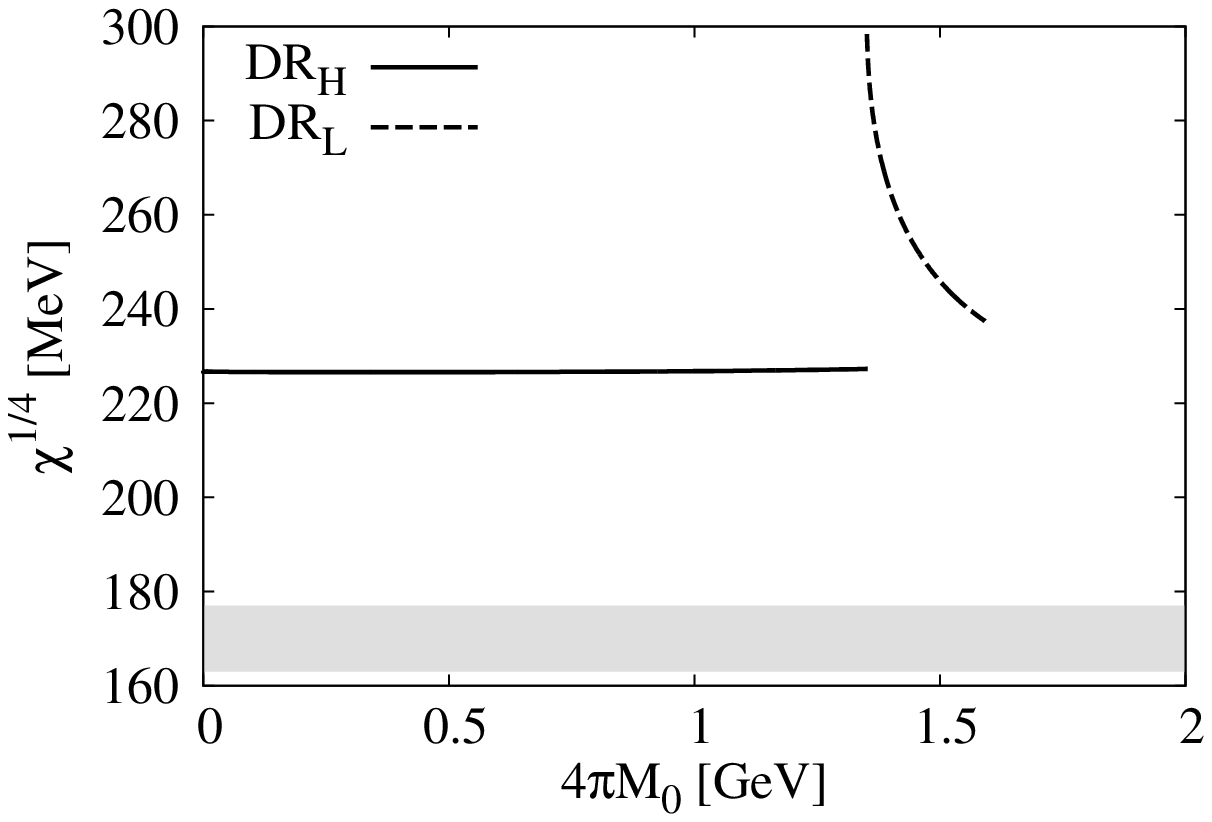}
   \caption{\label{fig:topo}
     $\chi^{1/4}$ in the 3D, 4D, PV, PT (left) and the DR (right).
     The gray areas are the range evaluated by the lattice QCD
     simulations~\cite{Alles:1996nm}.}
\end{center}
\end{figure}
Finally in Fig.~\ref{fig:topo}, we show the topological
susceptibility calculated through the topological charge density
presented in \ref{subsec:topo}.
We note that the result in the 3D case is the closest to the one by
the lattice QCD simulations, and the curves in the 4D, PV and PT enter
the consistent region with lattice QCD for high $\Lambda$, while the
DR does not touch the gray region. It is interesting to see that the
obtained curves show a certain similarity with the results of
$\eta$ mass; the 3D and 4D plots decrease for small $\Lambda$
then increase for large $\Lambda$ while the PV and PT cases go up
with increasing $\Lambda$. The reason may be originated from the
resembling systems of the equations for $m_\eta$ and $\chi$ where
the matrix form, in particular the nonzero off diagonal elements
due to the $U_{\rm A}(1)$ anomaly, plays a crucial role. The results
in the DR are almost constant around $\chi^{1/4} = 225$MeV which
is larger than the other regularization methods.
Since one extra parameter is introduced in DR, the $\eta$ meson 
mass is fixed as an input parameter.  If we accept some tolerance for
the $\eta$ meson mass which is used as an input parameter,
we can find a smaller topological susceptibility \cite{Inagaki:2010nb}.

\section{\label{sec:discussion}%
Discussions}
The paper has been devoted to the systematic analyses on the parameter
fitting and resulting predictions within various regularization procedures. 
We think now it may be intriguing that we give the detailed speculation on
the tendency of the obtained parameters and the model predictions.

\subsection{\label{subsec:high_L}%
High scale behavior}
Although the model is no longer effective for very high energy scale above
$\Lambda_{\rm QCD} \sim 1$GeV, we think that it is still worth studying how
the model behaves at high energy. Here we are going to present the analysis
on the asymptotic behavior of the current model. From the resulting parameters
evaluated in Fig.~\ref{fig:para}, we note the interesting feature that the
dimensionless coupling strengths $G\Lambda^2$ and $K\Lambda^5$ seem
to approach some constant values at large $\Lambda$ limit.
This can be understood by the discussions along Ref.~\cite{Inagaki:2013hya}.
For example, in the 3D case we read the dominant contribution
of $\phi_i^{\rm 3D} (=-i {\rm tr}S_i^{\rm 3D})$,
\begin{equation}
  \phi_u^{\rm 3D} \simeq -\frac{N_c}{2\pi^2}m_u^* \Lambda_{\rm 3D}^2,
\end{equation}
in the gap equation (3). Then we have
\begin{equation}
  1 \simeq \frac{N_c}{2\pi^2} \left( 4G \Lambda_{\rm 3D}^2 + 
     \frac{N_c}{\pi^2} K m_s^* \Lambda_{\rm 3D}^4 
    \right) .
\end{equation}
From Figs.~\ref{fig:para} and~\ref{fig:PM}, the relations,
$G \sim \Lambda^{-2}$, $K \sim \Lambda^{-5}$, and
$O(m_s^*/\Lambda) < 1$, can be read for high $\Lambda$, then we see

\begin{equation}
  G^{\rm 3D}\Lambda_{\rm 3D}^2 \simeq \frac{\pi^2}{2N_c} .
\end{equation}
In the same way, we have
\begin{equation}
  G^{\rm 4D}\Lambda_{\rm 4D}^2 \simeq \frac{\pi^2}{N_c} ,
\end{equation}
in the 4D case.
By noting that the PV and PT cases have the same asymptotic behavior with
the 4D methods, then we expect that the $G\Lambda^2$ approaches to
$1.64$ in the 3D case, and $3.29$ in the 4D, PV, and PT methods.  Our
numerical results are $G\Lambda^2=1.63$ for $\Lambda=8.1$GeV in the 3D,
$3.24$ for $\Lambda=8.3$GeV in the 4D, $3.16$ for $\Lambda=8.3$GeV
and $2.98$ for $\Lambda=2.0$GeV in the PT. These results seem to support the
above discussion. By following the similar analyses on the equation for the $\eta$
masses, we can also confirm the asymptotic behavior of $K$.
This is the numerical reason why the two coupling strengths approach to some
constant values at high energy limit.

\subsection{\label{subsec:onDR}%
Remarks on the dimensional regularization}
We have used the quantity $4\pi M_0$ as the counterpart of the cutoff scale
in the DR with fixed $m_\eta$. It may be worth reconsidering on whether the
current treatment is appropriate for the model analyses here. Since the DR
makes integrals finite by changing the integral kernel, not by restricting the
integration interval, the relation between these two scales are actually not
quite clear in this effective model approach.
The typical integrals go to infinity in the $D \to 4$ limit, so one might think it is
nice to treat $D \to 4$ limit as the counterpart of the $\Lambda \to \infty$
limit. However, the integral in the $D \to 4$ limit can be finite thanks to the
conditions in subsection \ref{subsec:procedure}.
The key point is the relationship between the mass scale parameter, $M_0$,
and $D$ which works so as to suppress the integral for higher dimensions
($M_0 \to 0$ for $D \to 4$) as studied in \cite{Inagaki:2013hya, Inagaki:2014kta}.
Due to this suppression, the resulting integrals, meaning the integrals with
factor $M_0^{4-D}$, do not necessarily increase with respect to $D$. 
Other correspondences between the cutoff and the dimensional regularizations
are discussed in Refs.~\cite{Krewald:1991tz} and \cite{Inagaki:1997}.
Based on the above discussion and the observation of the tendencies on the
obtained model parameters and the predicted quantities, we think that our present
treatment is effectively acceptable and the mass scale $4\pi M_0$ parameter 
can be considered as the quantity which closely relates to a cutoff scale.

\subsection{\label{subsec:system}%
System of parameter fitting}
As the final discussion on the parameter fitting, we present the fact that the
fitting, in particular, the existence of the parameters is sensitively determined
by the property of the $\eta^{\prime}$ mass equation.

\begin{figure}[h!]
\begin{center}
   \includegraphics[width=7.5cm,keepaspectratio]{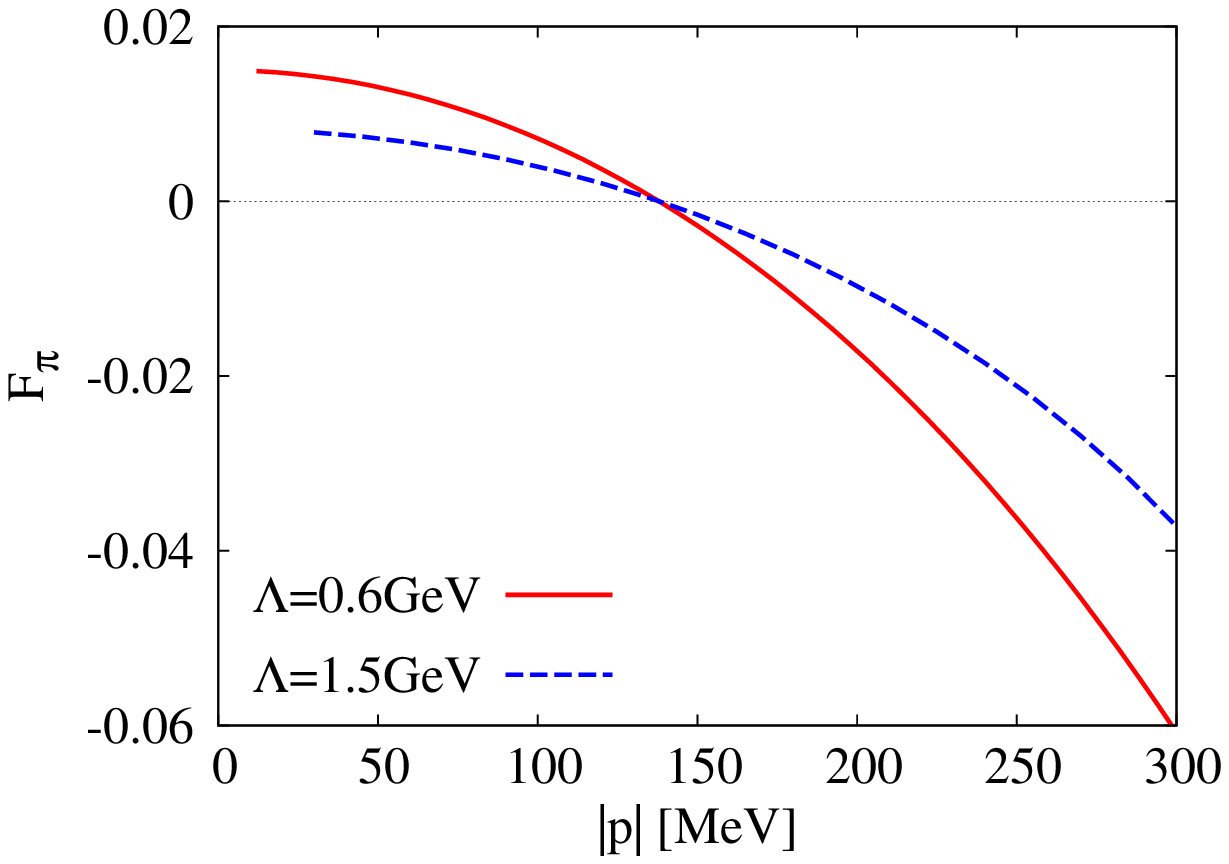}
   \includegraphics[width=7.5cm,keepaspectratio]{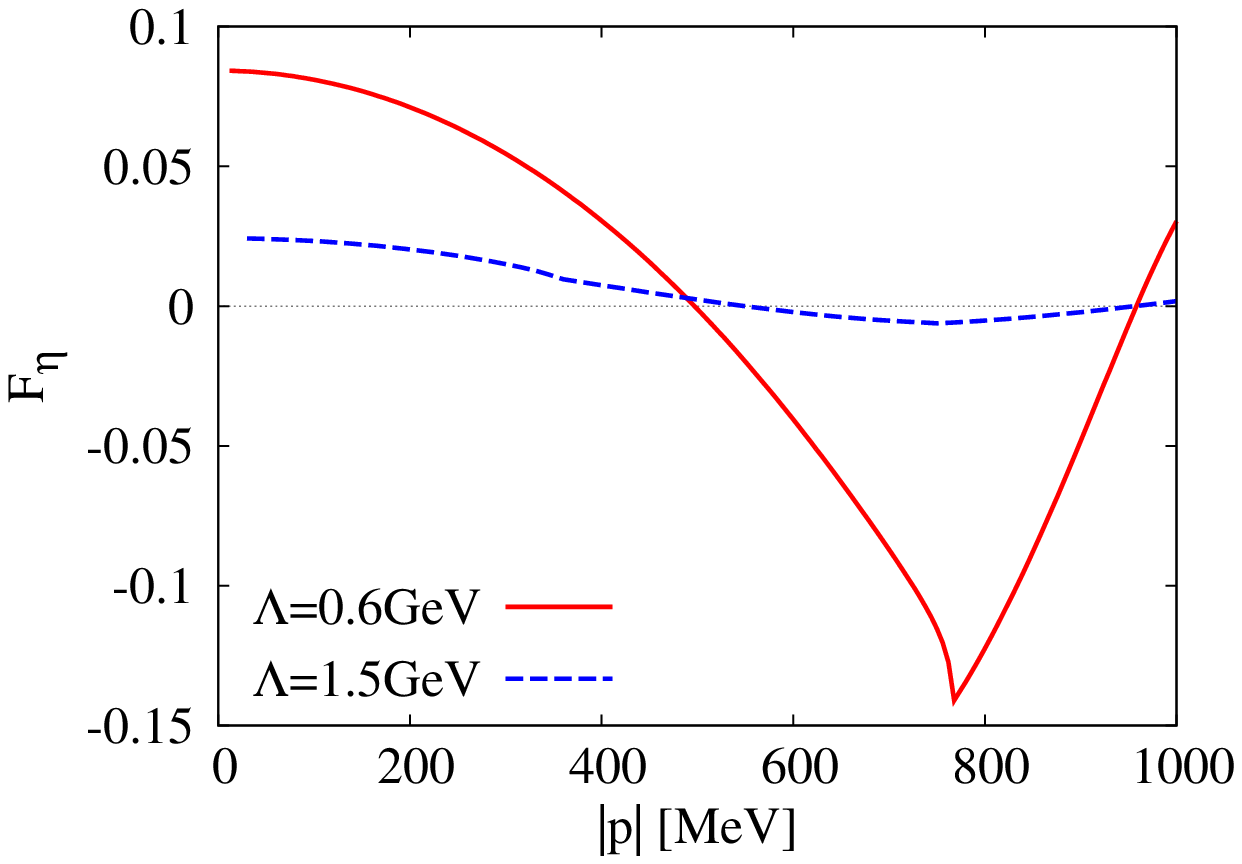}
   \caption{\label{fig:eq_meson}
        Typical behaviors of the equations for the pion (left)
        and $\eta^{\prime}$ (right) masses in the 3D.}
\end{center}
\end{figure}
Figure \ref{fig:eq_meson} displays the typical results on the functions for
the pion and $\eta$ masses; where the intersection points between
each curve and dashed line indicate the values of the meson masses,
$|p| =m_\pi(138$MeV) and $m_\eta^{\prime}(958$MeV).
Since the qualitative feature of the functions is similar in the other regularization
cases, we show the results of the 3D cutoff method as the representative figure
for the explanation on the parameter fitting. It is also worth mentioning that
the curve for the kaon equation is similar to the pion case. One notes that the
equations for the pion present monotonic decrease with respect to $|p|$,
while the result for the $\eta$ mass function exhibits some complex structure.
This comes from the determinantal form of the equation, which is due to the
non-vanishing off-diagonal elements in the $\eta-\eta^{\prime}$ system
as explicitly seen in the Appendix A.3.

The properties with changing $\Lambda$ on the functions are as follows.
Although the slope with respect to $|p|$, namely
$\partial \mathcal{F} / \partial |p|$, changes according to $\Lambda$ for
$\mathcal{F}_\pi$, we can always find the solution. On the other hand, the
change of the slope of $\mathcal{F}_{\eta}$ becomes important when we
try to find the solutions on it; the absolute value of the slope decreases
with increasing $\Lambda$ as seen from Fig. \ref{fig:eq_meson}, then the
range of $\mathcal{F}_{\eta}$ gets
smaller for high $\Lambda$, leading the numerical difficulty of searching
the solution. For small $\Lambda$, on the other hand, the absolute value
of the slope becomes larger, which leads the move of the curve and we
eventually reach the point where we can no longer find the solution for
$m_{\eta^{\prime}}=958$MeV at some low $\Lambda$. Thus the parameter
fitting is sensitively affected by the equation for determining the
$\eta^{\prime}$ mass.

\section{\label{sec:conclusion}%
Summary and conclusions}
In this paper we have applied various regularization procedures
to the NJL model then carried out the meticulous parameter fitting.
One of the main results of this paper is the parameters sets investigated
in the same conditions and the input quantities for each regularization.
Similar behavior is observed for the model parameters
in the 4D, PV and PT. 
The model parameters are determined even for extremely high scale
beyond the hadronic energy, which, we think, is surely interesting since
one can consider the ultraviolet region of the model. So we analyzed
the asymptotic behavior of the model through considering the
$\Lambda \to \infty$ limit. We then analyzed the asymptotic
behavior of the model through considering the $\Lambda \to \infty$
limit. Where we saw the coupling strengths in the dimensionless form,
$G\Lambda^2$ and $K\Lambda^5$, for the effective four-
and six-point interactions approach to constant values. This is as well
the interesting feature of the current model.

After setting the model parameters, we evaluated the predicted values,
$\phi_i, m^*_i, f_{\rm K}, m_\sigma$ and $\chi$ in each regularization.
We studied whether the predicted
quantities can indicate the values close to the experimental observations.
We found that the obtained physical predictions show satisfactory close
values to the empirical ones in the 3D, 4D, PV and PT methods, while
some quantity has different order in the DR. This may indicate that the
higher order corrections may be important due to the change of an
integral kernel with smaller spacetime dimensions in the dimensional
regularization. It should be noticed that one extra parameter is necessary
in DR case.  Thus we impose to generate the $\eta$ meson mass as input.
If we relax the condition, we can find a parameters set to show more
appropriate value for the other physical predictions.

We believe that the current analyses is useful for the readers
who want to investigate the properties of mesons and hadrons,
the transition phenomena of QCD such as the chiral phase transition,
and the calculation procedures in various regularizations. Also, we
think the obtained model parameters are useful since they enable
us to study a lot of physical quantities by using various regularization
methods. We plan to study the phase transition of the chiral symmetry
breaking by using the obtained parameters in various regularization
methods in future.

\section*{Acknowledgements}
HK is supported by MOST 103-2811-M-002-087.
TI is supported by JSPS KAKENHI Grant Number 26400250 and and 15H03663.

\appendix
\section{\label{app:meson}%
Meson properties and topological susceptibility}
Here we present the prescriptions on how one studies the physical observables
in the current model treatment. The masses of mesons are investigated by using
the Bethe-Salpeter equations which give the green functions for the composite
particles. Similarly, the decay constants for mesons and the topological susceptibility
are evaluated based on the diagrammatic calculations incorporating the constituent
quark propagator. The detailed derivations of the equations are presented in some
review papers, see for instance~\cite{Klevansky:1992qe}.

\subsection{\label{subsec:pi_K}%
Pion, sigma and kaon masses}
The masses of the pion, sigma and kaon are evaluated at the pole position
of the propagators derived from the random phase approximation with the
leading order of the $1/N_c$ expansion. The propagators for these mesons
are given by
\begin{eqnarray}
  \Delta_{\rm m}(p^2)
  \simeq \frac{2K_{\rm m}}{1-2  K_{\rm m} \Pi_{\rm m}(p^2)} 
  \simeq \frac{g_{{\rm m}qq}^2}{p^2 - m_{\rm m}^2},
  \label{BSE}
\end{eqnarray}
with the effective couplings for each channel (${\rm m}= \pi, \sigma$ and
${\rm K}$),
\begin{eqnarray}
 K_{\pi, \sigma} &=& G - \frac{1}{2} K \phi_s, \\
 K_{\rm K} &=& G - \frac{1}{2} K \phi_u,
\end{eqnarray}
and the quark-antiquark loop contributions $\Pi_{\rm m}(p^2)$ are given by
\begin{eqnarray}
  \Pi_{\pi}(p^2) = 2 \Pi_{\rm p}^{uu}(p^2), \\
  \Pi_{\sigma}(p^2) = 2 \Pi_{\rm s}^{uu}(p^2), \\
  \Pi_{\rm K}(p^2) = 2 \Pi_{\rm p}^{su}(p^2), 
\end{eqnarray}
for each meson with 
\begin{eqnarray}
  \Pi^{ij}_{\rm p}(p^2) &=& \int \frac{\md^4q}{i(2\pi)^4}
  {\rm tr}\bigl[ \gamma_5 S^i(q+p/2) \gamma_5 S^j(q-p/2)\bigr], \\
  \Pi^{ij}_{\rm s}(p^2) &=& \int \frac{\md^4q}{i(2\pi)^4}
  {\rm tr}\bigl[ S^i(q+p/2) S^j(q-p/2)\bigr],
\end{eqnarray}
where the suffices ${\rm p}$ and ${\rm s}$ indicate the pseudo-scalar and scalar
channels, respectively. One obtains the following expressions after a bit of algebra 
\begin{eqnarray}
  \Pi_{\rm p}^{ij} (p^2)
  &=&
   \frac{  i{\rm tr} S^i }{2m^*_i} 
  +\frac{  i{\rm tr} S^j }{2m^*_j}
  +\frac{1}{2} \left[ p^2-(m^*_i-m^*_j)^2 \right]  I_{ij}(p^2) , \\
  \Pi_{\rm s}^{ij} (p^2)
  &=&
     \frac{ i{\rm tr} S^{i} }{2m^*_i} 
    +\frac{ i{\rm tr} S^{j} }{2m^*_j} 
    +\frac{1}{2} \left[ p^2- (m^{*}_i + m^*_j )^2 \right]  I_{ij}(p^2) .
  \label{eq:scalar}
\end{eqnarray}

Since the following relations should hold
\begin{equation}
 1-2K_{\rm m} \Pi_{\rm m}(p^2) |_{p^2=m_{\rm m}^2} = 0,
\label{eq:meson_mass}
\end{equation}
at the pole position, then the meson mass $m_{\rm m}$ is given by
Eq.~(\ref{eq:meson_mass}).

\subsection{\label{subsec:pi_decay}%
Pion and kaon decay constants}
The pion and kaon decay constant are calculated through evaluating the following
one-loop amplitude,
\begin{eqnarray}
  i p^{\mu} f_{\rm m} \delta^{\alpha \beta}
  =
  \langle 0| \bar{q} \frac{T^\alpha}{2} \gamma^{\mu}
             \gamma_5 q  |{\rm m}^\beta \rangle
 = - \sum_{kl} \int  \frac{\md^4 q}{(2\pi)^4}
  {\rm tr}\left[\gamma^\mu \gamma_5 \frac{T^\alpha_{k l}}{2} 
  S_l (q+p/2) g_{\rm m qq} \gamma_5 T^{\beta \dagger}_{l k} S_k (q-p/2) \right],
\end{eqnarray}
with $T^\alpha = (\lambda_1 \pm i \lambda_2)/\sqrt{2}$ for $\pi^{\pm}$
channels and $T^\alpha = (\lambda_4 \pm i \lambda_5)/\sqrt{2}$ for $K^{\pm}$
channels,
and the coupling strengths for the meson-quark-quark interaction,
$g_{{\rm m} qq}$, evaluated by
\begin{equation}
  g_{\rm m qq}^2 (p^2) =
      \left( \frac{\partial \Pi_{{\rm m}} }{\partial p^2} \right)^{-1}.
\end{equation}
After some algebra, we have
\begin{eqnarray}
 f_\pi &=& m^*_u g_{\pi {\rm qq}}(0)  I_{uu} (0) , 
\label{eq:pi_decay} \\
 f_{\rm K} &=& g_{\rm K qq}(0) \left[ m^*_u I_{us}(0) + (m^*_s - m^*_u) 
     \int_0^1 \md x \int \frac{\md^4 q}{i(2\pi)^4} {\rm tr}
     \frac{x}{\{q^2 - \Delta_{us}(0) + i \epsilon\}^2} \right] ,
\label{eq:K_decay}
\end{eqnarray}
where
equations are evaluated at $p^2=0$ and give the pion and
kaon decay constants, respectively~\cite{Klevansky:1992qe}.

\subsection{\label{subsec:eta_mass}%
$\eta$ and $\eta^{\prime}$ masses}
Compared with the pion and kaon masses,
there appears the complexity for the
$\eta$--$\eta^{\prime}$ system, where the propagator can be
written by the matrix form as
\begin{equation}
  \hat{\Delta}_{\eta} (p^2) 
  =  2\hat{\bf K} \left[ 1-2 \hat{\Pi}  (p^2) \hat{\bf K}  \right]^{-1}, 
\label{}
\end{equation}
where  $\hat{\bf K}$ and $\hat{\Pi}$ represent $2\times 2$
matrices whose elements are given by
\begin{eqnarray}
K_{00} &=& G + \frac{1}{3} K(2 \phi_u + \phi_s), \\
K_{88} &=& G - \frac{1}{6} K(4 \phi_u - \phi_s), \\
K_{08} &=& K_{80}= -\frac{\sqrt{2}}{6} 
                  K(   \phi_u - \phi_s),\label{K_08}\\
\Pi_{00} &=& \frac{2}{3}
     \left[ 2 \Pi_{\rm p}^{uu}(p^2) +\Pi_{\rm p}^{ss}(p^2) \right], \\
\Pi_{88} &=& \frac{2}{3}
     \left[ \Pi_{\rm p}^{uu}(p^2) 
             +2\Pi_{\rm p}^{ss}(p^2) \right], \\
\Pi_{08} &=& \Pi_{80}=\frac{2\sqrt{2}}{3} 
     \left[  \Pi_{\rm p}^{uu}(p^2) 
                 - \Pi_{\rm p}^{ss}(p^2) \right].
\label{Pi_08}
\end{eqnarray}
Then the condition which determines the $\eta$ and $\eta^{\prime}$ masses
becomes
\begin{equation}
  \det \left[ 1-2 \hat{\Pi}  (p^2)  \hat{\bf K}  \right]
  \Bigl|_{p^2=m_{\rm m}^2}  = 0.
\label{eq:eta}
\end{equation}
The explicit expressions for the numerical calculations are shown
in~\cite{Inagaki:2010nb}.

\subsection{\label{subsec:topo}%
Topological susceptibility}
The topological susceptibility,
\begin{equation}
\chi = \int \md^4x \,
  \langle 0| TQ(x)Q(0) |0\rangle_{\rm connected},
\end{equation}
is calculated from the following topological charge density~\cite{Hatsuda:1994pi},
\begin{equation}
  Q(x)\equiv\frac{g^2}{32\pi^2}F^a_{\mu\nu} \tilde{F}^{a\mu\nu}
  = 2K \, {\rm Im} \left[ \det \bar{q}(1-\gamma_5)q \right].
\end{equation}
The explicit formula is evaluated in \cite{Fukushima:2001hr}, which reads
\begin{eqnarray}
  \chi
  &=& -4K^2
    \phi_u^2 \left[ \phi_u \phi_s
    \left(\frac{2\phi_s }{m_u^*} + 
    \frac{\phi_u}{m_s^*} \right) \right.
    \nonumber \\
  &&+ \left\{ \frac{1}{\sqrt{6}} (2\phi_s + \phi_u)
    \bigl( \Pi_{00}(0), \Pi_{08}(0) \bigr) \right.
   + \left. \frac{1}{\sqrt{3}} (\phi_s - \phi_u)
    \bigl( \Pi_{08}(0), \Pi_{88}(0) \bigr)
    \right\} \Delta^+(0)
    \nonumber \\
  && \times\left\{ \frac{1}{\sqrt{6}} (2 \phi_s + \phi_u)
    \left(
    \begin{array}{c}
       \Pi_{00}(0) \\
       \Pi_{08}(0) 
    \end{array}\right) \right.
   + \left. \frac{1}{\sqrt{3}} (\phi_s - \phi_u)
    \left( \left.
    \begin{array}{c}
     \Pi_{08}(0) \\
     \Pi_{88}(0)
    \end{array} \right) \right\} \right] .
\label{chi}
\end{eqnarray}






\end{document}